\documentclass{jfm}
\usepackage{caption}
\usepackage{subcaption}
\usepackage{graphicx}
\usepackage{epstopdf, epsfig}
\usepackage{bbold}
\usepackage{siunitx}
\usepackage{cmll}
\usepackage[mathcal]{eucal}
\usepackage{ucs}
\usepackage[utf8x]{inputenc}
\usepackage{amsmath}
\def\bm#1{\mbox{\boldmath{$#1$}}}
\def\rr#1{(\ref{#1})}

\def\cprime{$'$}
\newcommand{\be}{\begin{equation}}
\newcommand{\ee}{\end{equation}}

\graphicspath{ {./Production_files_inertial_wavesI} }

\shorttitle{Taylor columns and inertial-like waves in odd viscous liquids}
\title{Taylor columns and inertial-like waves in a three-dimensional odd viscous liquid}
\author{E.Kirkinis%\aff{1}
\corresp{\email{kirkinis@northwestern.edu} }
\and M. Olvera de la Cruz }
\affiliation{Department of Materials Science \& Engineering, Robert R. McCormick School of Engineering and Applied Science, Northwestern University, Evanston IL 60208 USA\\ Center for Computation and Theory of Soft Materials, Northwestern University, Evanston IL 60208 USA
}

\begin{document}

\maketitle

\begin{abstract}
Odd viscous liquids are endowed with an intrinsic mechanism that tends to restore a displaced particle
back to its original position. Since the odd viscous stress does not dissipate energy, inertial oscillations 
and inertial-like waves can become prominent in such a liquid. In this article we show that an odd viscous liquid 
in \emph{three} dimensions
gives rise to such axially symmetric waves and also to plane-polarized waves. We tacitly assume that an
anisotropy axis giving rise to odd viscous effects has already been established and proceed to investigate
the effects of odd viscosity on fluid flow behavior. 
Numerical simulations of the full Navier-Stokes equations show the existence of inertial-like
waves downstream a body that moves slowly along the axis of an odd viscous liquid-filled cylinder.  
The wavelength of the numerically-determined oscillations agrees well with the developed theoretical framework.
When odd viscosity is the dominant effect in steady motions,  
a modified Taylor-Proudman theorem leads to the existence of Taylor columns inside such a liquid. 
Formation of the Taylor column can be 
understood as a consequence of helicity segregation and energy transfer along the cylinder axis at group velocity, by the accompanying
inertial waves, whenever the reflection symmetry of the system is lost.  
A number of Taylor column characteristics known from rigidly-rotating liquids, 
are recovered here for a \emph{non-rotating} odd viscous liquid. These include counter-rotating 
swirling liquid flow above and below a body moving slowly along the anisotropy axis. 
Thus, in steady motions, odd viscosity acts to suppress variations of liquid velocity in a direction parallel to the anisotropy axis, inhibiting vortex stretching and vortex twisting. 
In unsteady and nonlinear motions odd viscosity enhances the vorticity along the same axis, thus affecting both vortex stretching and vortex twisting.  
\end{abstract}

%--------------------------------------------------------------------------
% Maple calculations are in Andreev/Noncentrosymmetric/axial_waves_mw
%---------------------------------------------------------------------------
\section{Introduction}
{\citet{Avron1995} showed that the viscosity tensor $\eta_{\alpha \beta \gamma \delta}$ of the 
Cauchy stress of a classical liquid 
\be \label{sabcd}
\sigma_{\alpha\beta} = \eta_{\alpha \beta \gamma \delta} V_{\gamma \delta}
\ee
can be decomposed into a symmetric and an antisymmetric part
: $\eta_{\alpha \beta \gamma \delta}=\eta^S_{\alpha \beta \gamma \delta} + \eta^A_{\alpha \beta \gamma \delta}$ where 
\be
\eta_{\alpha \beta \gamma \delta}^S = \eta^S_{\gamma \delta\alpha \beta }
\quad 
\textrm{and}
\quad 
\eta^A_{\alpha \beta \gamma \delta} = -\eta^A_{\gamma \delta\alpha \beta }, 
\ee
where $V_{\gamma\delta} = \frac{1}{2}\left( \frac{\partial u_\gamma}{\partial x_\delta} + \frac{\partial u_\delta}{\partial x_\gamma}\right)$ is the rate-of-strain tensor and $u_\alpha$ the liquid velocity. 
The stress tensor \rr{sabcd} based on $\eta^S$ is dissipative. 
It results in viscous heating \citep[\S 49]{Landau1987} which is a positive-definite quadratic form 
$\textrm{tr}(\sigma V) = V_{\alpha\beta}\eta^S_{\alpha \beta \gamma \delta}V_{\gamma\delta}>0$.  
$\eta^A$ does not contribute to viscous heating due to its antisymmetry between the first and the second pairs of
indices. 

The non-dissipative stress $\eta^A_{\alpha \beta \gamma \delta} V_{\gamma \delta}$ is called the odd viscous (or anomalous) stress tensor and its accompanying coefficients odd or Hall viscosity coefficients. This type of behavior can be induced, for instance, by a magnetic field, giving rise to an
\emph{anisotropy axis} along its direction. 
\citet{Avron1995} provided a clear physical interpretation of the odd stress tensor that carries over
to classical systems. Compression or dilatation gives rise to shear and vice-versa. Thus, it is easy to 
show, for instance, that a cylinder rotating about its axis in an odd viscous liquid gives rise to 
a stress $\sigma_{rr}^o %\equiv -\eta_o (\partial_r v_\phi - \frac{1}{r} v_\phi) 
= 2\eta_o\Omega$ directed normal to cylinder surface \citep{Avron1998,Kirkinis2023null}, $\eta_o$ is the dynamic coefficient of odd viscosity and $\Omega$ the constant angular velocity of the cylinder.

The ramifications of odd viscosity in classical mechanical systems have been investigated only in the 
recent literature. Noteworthy are experiments showing 
blobs of a liquid composed of micron-size spinning magnets whose surface undulations were 
attenuated by a shear stress-induced odd normal stress (the shear stress is generated by the collective
rotation of the particles close to the free surface, cf. \citep{Soni2019}). In three dimensions \citet{Khain2022}  
showed that odd viscous liquids, in the absence of inertia, give rise to 
unconventional fluid-flow behavior such as the stabilization of a sedimenting cloud of particles due to an 
odd viscosity-induced azimuthal velocity field generated by the gravitational stretching of the cloud
in the axial direction, in agreement with the physical interpretation of \citet{Avron1995}. In two dimensions
such an odd viscosity-induced azimuthal component can be seen in the radial expansion
of a bubble \citep{Ganeshan2017}. 
From a microscopic point of view, the mechanisms that may give rise to an odd viscosity coefficient 
include broken parity, broken time-reversal and microscopic torques. The latter can be traced back 
to the literature of liquids endowed with rotational degrees of freedom \citep{Dahler1961} which 
have been employed in the continuum description of magnetic liquids \citep{Rinaldi2002Thesis,Kirkinis2017}.  
Other odd viscosity-induced phenomena have been succinctly collected in the recent review by \citet[Fig. 1]{Fruchart2023}.

The stabilizing behavior induced by odd viscosity in \citep{Soni2019} as well as in other references that have appeared in the recent literature (see e.g. \citep{Kirkinis2019b}), implies that
the odd viscous stress may endow its medium with an intrinsic restoring property. }
Since the odd viscous stress
does not dissipate energy, an excitation given to the fluid may establish an oscillation. 
Such an oscillation may further initiate wave propagation and periodic expansion and contraction
in a plane perpendicular to the propagation direction. This type of motion (in a non-odd viscous liquid)  is called an inertial wave
and is present in the ocean
driving its upper mixing \citep{Asselin2020}, constituting half of its energy
and being responsible for the majority of its vertical shear.  Inertial waves also appear in the celestial sphere \citep{Ogilvie2013} 
and in the technology of propulsion \citep{Gao2020}. It has also been argued that inertial waves  
in the earth's interior are associated with a dynamo dipole and with helicity segregation \citep{Davidson2014,Davidson2018}. 
{Waves in odd viscous liquids have been investigated in a number
of occasions. These include gravity waves \citep{Abanov2018} in incompressible liquids, shock waves \citep{Banerjee2017} and topological sound waves propagating along an interface
\citep{Souslov2019} in two-dimensional compressible liquids. 
Among other reasons, the topological waves are interesting because they are reminiscent of zonal currents associated with wall modes 
in rapidly-rotating Rayleigh-B\'enard convection, although the latter are three-dimensional 
and are driven by thermal forcing \citep{Knobloch2022}.
}

\begin{figure}
\vspace{5pt}
\begin{center}
\includegraphics[height=2in,width=5in]{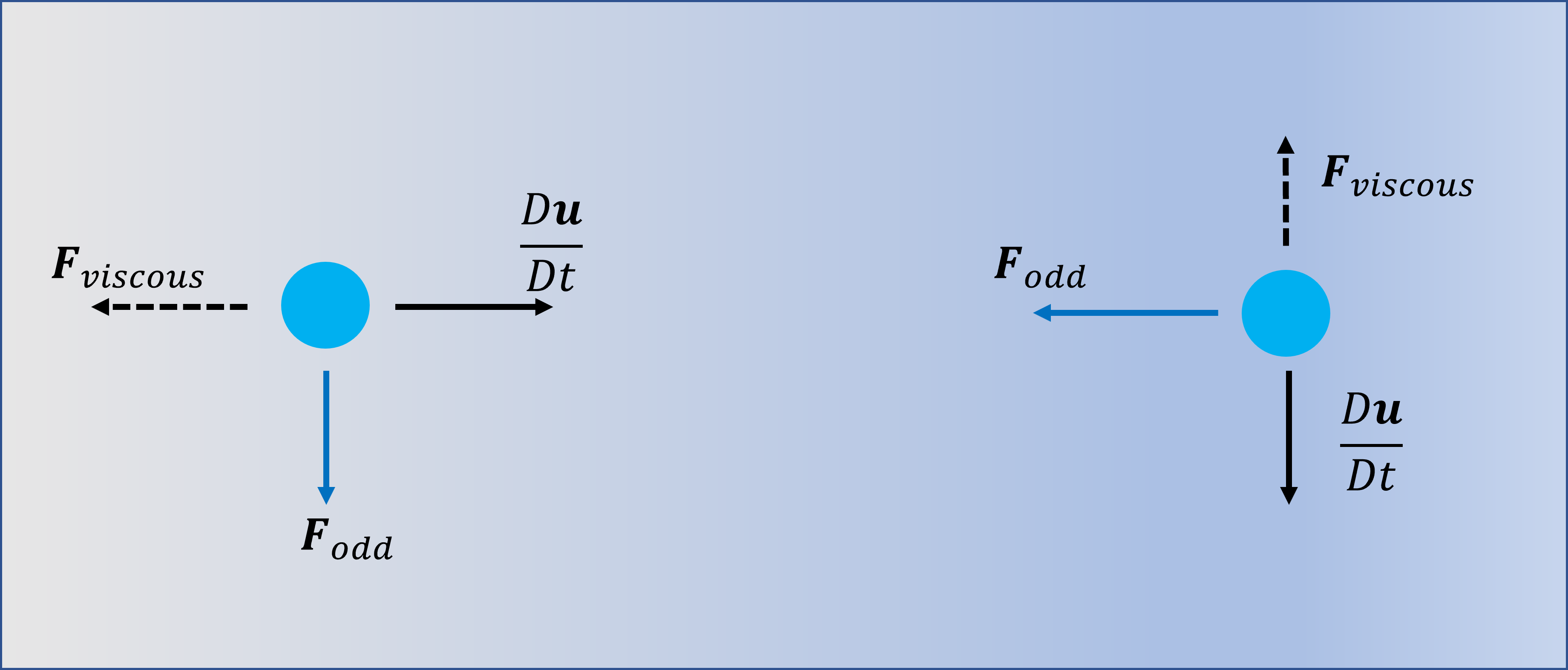}
\vspace{-0pt}
\end{center}
\caption{Restoring mechanism associated with odd viscosity in two dimensions. Left: acceleration of a fluid particle is resisted by the viscous force; its odd viscous counterpart
acts perpendicularly to the axis of the viscous force. Right: The new tangential motion acquired by the fluid particle is
resisted by the viscous force and thus a new odd viscous force acts perpendicularly. The direction
of the latter is \emph{opposite} to the original fluid-particle acceleration. 
\label{fig1}  }
\vspace{-0pt}
\end{figure}

{A three-dimensional odd viscous liquid may also give rise to Taylor columns. 
Taylor columns are known to form when bodies move slowly in (non-odd viscous) rapidly rotating liquids
\citep{Davidson2013book}. For instance, a slowly-moving body along the axis of a rigidly-rotating liquid
gives rise to a flow whose component along the axis of rotation can decouple from its lateral plane counterpart} (by lateral plane we will
mean the plane whose normal is the anisotropy axis). Thus, a Taylor column will form whose speed
will be identical to the speed of the slowly-moving body
it circumscribes. 
{There are certain restrictive conditions that need to be satisfied in order for a 
Taylor column to form. These are the small Rossby and Ekman numbers as are defined in 
Eq. \rr{RoEk} which thus require high angular velocity of rotation (and slow motions in the rotating frame)
and low values of the shear viscosity coefficient. }
Taylor columns are present in a multitude of diverse areas: cold water and low
salinity domes form over seamounts (eg. the Rockall, Faroe and 
Hutton Banks) of high chlorophyll and nutrient levels enabling larval diversity hotspots \citep{Dransfeld2009} by entraping plankton (presumably by entraining matter through Stewartson layers). 
A turbulent ocean below the icy crust of Enceladus and 
Europa is likely to transfer energy through counter-rotating zonal jets inside a Taylor column
\citep{Bire2022}.

{Taylor columns form in (non-odd viscous) rotating liquids because the Coriolis force is always perpendicular
to the velocity field (the latter is expressed in the frame of reference rotating with the liquid). Thus, a radial motion of a fluid particle
gives rise to a commensurate azimuthal component and vice-versa. This is reminiscent of the 
physical interpretation given by \citet{Avron1995} to odd viscosity and thus leads to the possibility of 
observing Taylor columns in such a liquid. 
As is the case in their rigidly-rotating counterparts, Taylor columns in (non-rotating) odd viscous liquids can 
be observed when certain restrictions are satisfied. These correspond to the smallness of the 
parameter
$ \mathcal{M}^{-1} = \frac{a U}{\nu_0}$ which requires a large coefficient of kinematic 
odd viscosity $\nu_o$ and slow motions ($U$ can be understood as the velocity of a body moving slowly
in an odd viscous liquid) and $a$ is a characteristic length-scale. In addition the effect of shear viscosity $\nu_e$ 
should be small and so should be the Ekman number $\mathcal{T}^{-1} = \frac{\nu_e}{\nu_o}$, 
see Eq. \rr{MaxworthyM}.  
}

{In Fig. \ref{fig1} we schematically demonstrate the restoring property induced by odd viscosity on a two-dimensional liquid. 
In three dimensions, when odd viscosity dominates over its shear counterpart, the restoring behavior can be established \emph{qualitatively} by balancing the inertial terms
in the Navier-Stokes equations with the odd viscous term (cf. \cite[\S 16.6]{Tritton1988} for the case of a rotating liquid).} 
Let $v>0$ be the velocity of an isolated unit mass fluid particle in a plane perpendicular to the 
anisotropy axis (say, in the azimuthal direction). Balance between inertia and odd viscous terms leads to 
\be \label{v2r}
\frac{v^2}{r}  = \nu_o \frac{v}{r^2}
\ee
giving
$v = \frac{\nu_o}{r}$, where $\nu_o = \eta_o/\rho>0$ is the odd kinematic coefficient of viscosity. Thus, the particle will move in circles of radius
$
r = \frac{\nu_o}{v
}.
$
The period $T_o$ of this rotation depends on the distance from the axis of anisotropy: $T_o = \frac{2\pi r^2}{\nu_o}$ and this shows that an odd viscous liquid is endowed with an intrinsic frequency
\be \label{omega0}
\omega_o = \frac{\nu_o}{r^2}. 
\ee
Although the above argument is only schematic, the derived expression
for the frequency $\omega_o$ \emph{is} recovered in the following quantitative analysis (see Eq. \rr{omega}), at least in the short
wave-length limit. 
A more satisfying argument supporting the restoring effect of odd viscous liquids is delegated to
the end of section \ref{sec: inertial1}.

{ In a three dimensional odd viscous liquid we observe waves whose motion resembles the inertial waves
occurring in (non-odd viscous) rigidly-rotating liquids. The anisotropy axis 
inherent in the odd stress plays the role of the rotating axis of a rigidly-rotating liquid. 
When the inertial-like waves induced by odd viscosity} are plane-polarized, there is a superposition of two oppositely directed waves,
particle paths are helical and this is reflected 
in the sign of the helicity density (vorticity times velocity) associated with each direction. 
When the propagation direction is at right angles to the odd anisotropy axis, the phase velocity vanishes 
and the system 
suffers a loss of reflection symmetry \citep{Moffatt1970}. Energy then propagates along the anisotropy axis
at maximum group velocity and is accompanied by helicity of a commensurate sign. This effect is called
``segregation of helicity'' in (non-odd viscous) rigidly-rotating liquids \citep{Davidson2018} and is believed to be important
in understanding the dipolar nature of planetary dynamos since in a planet,
mean helicity is spatially segregated having opposite signs at the 
northern and southern hemispheres, respectively.  

In this article we will tacitly assume that such an axis of anisotropy has already been established
and proceed by examining the consequences of the resulting odd viscous stress to fluid motions. 
Here we consider fluid motions in a three-dimensional odd viscous liquid. 
This paper proceeds in the following manner. In section \ref{sec: hydrodynamics} we describe the 
constitutive law for the odd viscous liquid that will give rise to the inertial-like waves \citep[\S13]{Landau1981}. 
Section \ref{sec: inertial1} establishes the existence of the inertial-like waves in an odd viscous liquid. These are waves that
propagate along the axis of anisotropy and wrapped in coaxial cylinders where liquid does not cross. 
We theoretically determine the frequency and 
wavelength of propagated modes. We provide a more satisfying, yet still qualitative discussion
of the ``elasticity'' of an odd viscous liquid. 
In section \ref{sec: inertial} we numerically solve the full Navier-Stokes 
equations for the slow motion of a sphere inside an odd viscous liquid. Such motions generate
liquid oscillations downstream the body. Their wavelength is in astonishing agreement to the 
theoretical value obtained from the inertial theory of section \ref{sec: inertial1}. 
In section \ref{sec: exterior} we establish the existence of inertial-like waves exterior to a cylinder
and extending to infinity. In section \ref{sec: plane} we derive the frequency, phase and group velocities
of three-dimensional plane-polarized waves. These differ from their axisymmetric counterparts 
derived in section \ref{sec: inertial1} which also vary in the propagation direction. They are also 
special as they segregate helicity and this is discussed in section \ref{sec: helicity}. 
In section \ref{sec: modified} we derive a modified Taylor-Proudman theorem. This means
that when odd viscosity dominates over shear viscosity and inertial terms, the motion in the lateral 
plane becomes decoupled to the motion of the fluid along the anisotropy axis. This opens up
the prospect of existence of Taylor columns in odd viscous liquids, which we explore in 
section \ref{sec: Taylor}. Since the study of Taylor columns entails overwhelming details, in order to provide some structure in our discussion we follow the map
set up by 
\citet{Maxworthy1970} for the slow motion of a particle. We thus solve numerically the Navier-Stokes
equations and find many similarities
to Maxworthy's work: counterrotating swirling motion above and below the sphere, a forward and a rearward
``slug'', 
a stagnant region, indication of an Ekman layer surrounding the sphere etc.

{In section \ref{sec: eta4} we revisit the foregoing results by introducing another part of the odd
stress tensor (the $\eta_4$ part of the stress in the notation followed by \citet[\S13 \& \S58]{Landau1981}).
This provides a complete picture of odd viscous effects that may be present in a liquid. A consequence 
of including both viscosity coefficients is the more diverse behavior displayed by the velocity field (it can now
resemble Kelvin functions). Helicity is still conserved in such a composite liquid when its velocity field is determined by
plane polarized waves. Throughout this paper we compare our results to those of the recent odd viscous liquid
literature. Thus, in a dedicated section \ref{sec: comparison} we summarize the results of these comparisons. }

{In Appendix \ref{sec: oddstress} we formulate the odd viscous stress tensor for a three-dimensional liquid in terms of the rate-of-strain tensor, in a manner that is common
to the continuum mechanics literature, and thus a compact
relation for the full stress incorporating both even (shear) and odd stresses is obtained. This
leads directly to the vanishing of the odd stress contribution in viscous heating. Hence, the constitutive law employed in this article represents a non-dissipative liquid.}  
The problems we discuss in this paper
present many similarities to flows generated in a rotating liquid and described in a rotating frame 
of reference. Thus, throughout the paper, where appropriate, 
we establish connections to these effects. We conclude in Appendix \ref{sec: rotating} by outlining a few facts
about rotating fluids that are of relevance to this article (although this article is not about rotating liquids).

\section{\label{sec: hydrodynamics}Constitutive relations of a three-dimensional odd viscous liquid}
{In fluid mechanics the constitutive law (the Cauchy stress tensor) of a \emph{Newtonian} liquid is 
usually introduced following the phenomenological approach, cf. \citep[\S 3.3]{Batchelor1967}
or more rigorously by employing the principle of objectivity cf. \citep{Truesdell1992}.
It is however possible to also introduce the notion of stress through the Onsager principle of the 
symmetry of the kinetic coefficients \citep[\S 13]{Landau1981}. 
When these 
coefficients (here the viscosity tensor $\eta_{\alpha\beta\gamma\delta}$ we introduced in Eq. \rr{sabcd}) 
depend on external fields, say $\mathbf{b}$, that 
change sign under time-reversal, the symmetry of the kinetic
coefficients is ensured when
% \be
% \gamma_{ik}(\mathbf{b}) = \gamma_{ki}(-\mathbf{b}).
% \ee
% When the coefficients depend on odd powers of the field $\mathbf{b}$,  the latter expression becomes
% $ - \gamma_{ki}(\mathbf{b})$, 
\be
\eta_{\alpha\beta\gamma\delta}(\mathbf{b})
= \eta_{\gamma\delta\alpha\beta}(-\mathbf{b}).
\ee
For an incompressible liquid the stress tensor \rr{sabcd}, subject to such a field, obtains the form }
%---------------------------------------------------------------------------------------------------------------------------------
%the stress tensor was calculated in Andreev/Hall viscosity/odd_viscosity_stress_components. Note that in the present 
% problem the sigma^o_xy component is u_x-v_y and this does not simplify to 2u_r as in the Null paper because
% here the problem is three dimensional
%---------------------------------------------------------------------------------------------------------------------------------
\begin{eqnarray}
\sigma_{\alpha\beta}' =2V_{\alpha\beta}(\eta + \eta_1)+ V_{\beta\gamma} \left[ 2(\eta_2 - \eta_1) b_\gamma b_\alpha + \eta_3 b_{\alpha\gamma}\right]
+ V_{\alpha\gamma} \left[ 2(\eta_2 - \eta_1) b_\gamma b_\beta + \eta_3 b_{\beta\gamma}\right] \nonumber &&\\
+ V_{\gamma\delta} 
\left[(\eta_1 + \zeta_1) \delta_{\alpha\beta}b_\gamma b_\delta + (\eta_1 - 4\eta_2)b_\alpha b_\beta b_\gamma b_\delta  + (2\eta_4-\eta_3)(b_{\alpha\gamma} b_\beta b_\delta + b_{\beta\gamma} b_\alpha b_\delta) \right], &&\label{seta1234} 
\end{eqnarray}
where $b_{\alpha\beta} = \epsilon_{\alpha\beta\gamma}b_\gamma$, $\eta_i,\zeta_i$ are viscosity coefficients, $V_{\gamma\delta} = \frac{1}{2}\left( \frac{\partial u_\gamma}{\partial x_\delta} + \frac{\partial u_\delta}{\partial x_\gamma}\right)$ and $\epsilon_{\alpha\beta\gamma}$ is the alternating
tensor.  

\begin{figure}
\vspace{-5pt}
\begin{center}
\includegraphics[height=2.5in,width=2.2in]{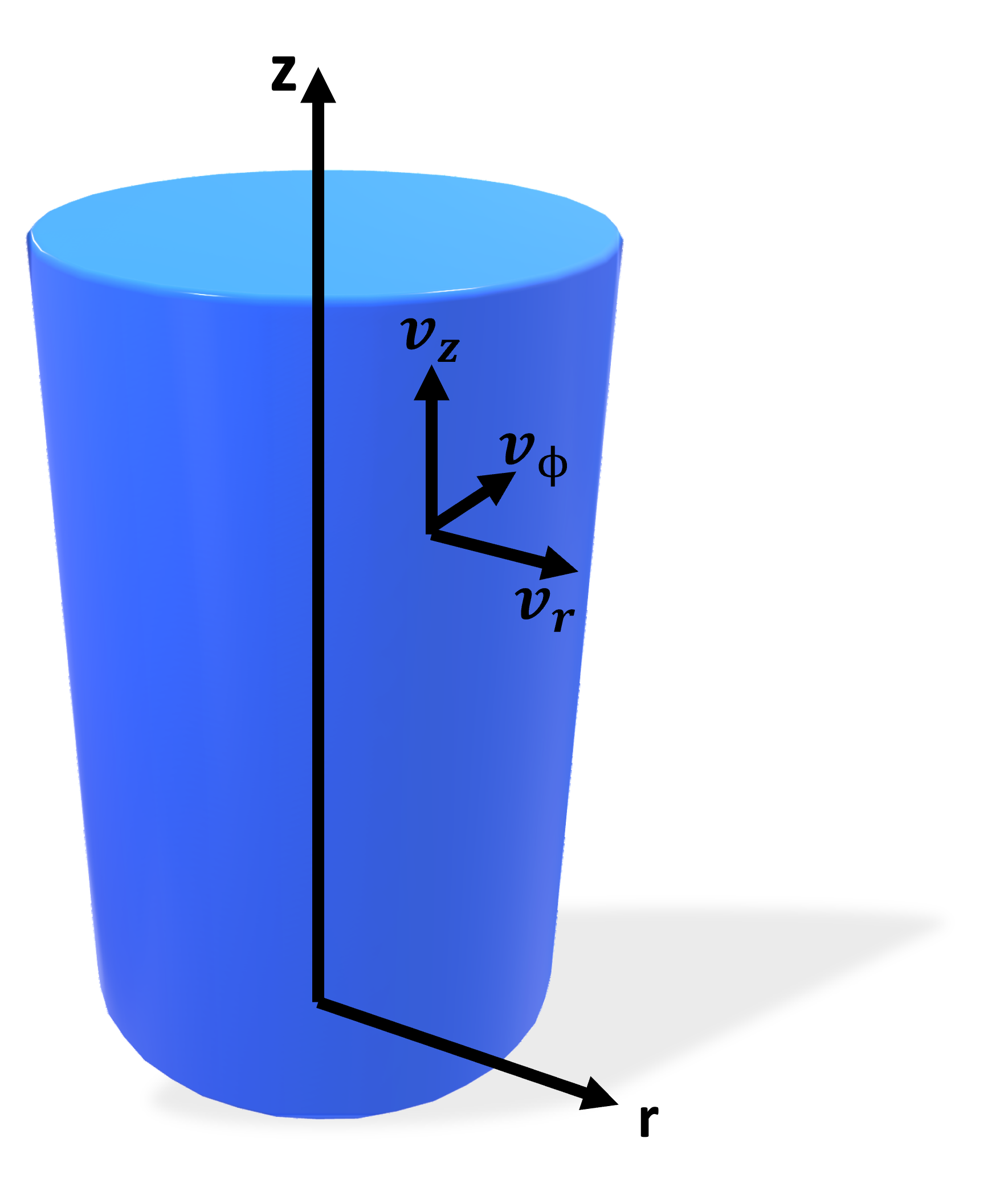}%{odd_axial_psi_zr}
\vspace{-0pt}
\end{center}
\caption{Three-dimensional odd viscous liquid in cylindrical coordinates with velocity field 
$\mathbf{v} = v_r \hat{\mathbf{r}} + v_\phi \hat{\bm{\phi}} + v_z \hat{\mathbf{z}}$. 
\label{cylinder}  }
\vspace{-0pt}
\end{figure}

{The physical system considered in this paper consists of an odd viscous liquid endowed with
odd coefficients $\eta_3$ and $\eta_4$ in \rr{seta1234} and the field $\mathbf{b}$ to lie in the $z$-direction, so $\mathbf{b} = \hat{\mathbf{z}}$.  

The presentation becomes opaque when both coefficients are employed simultaneously. 
We thus consider 
each one in turn}. Considering only $\eta_3 \neq 0$, 
we set $\eta = \eta_4=0$ and $\eta_2 =\eta_1 = - \zeta_1$. The corresponding odd stress tensor in polar cylindrical coordinates reads (cf. Fig. \ref{cylinder})
\be \label{sigma0}
\bm{\sigma}' = \eta_o 
\left(\begin{array}{ccc}
-\left(\partial_r v_\phi - \frac{1}{r}v_\phi + \frac{1}{r}\partial_\phi v_r \right) 
&  \partial_r v_r - \frac{1}{r}v_r - \frac{1}{r}\partial_\phi v_\phi & 0\\
\partial_r v_r - \frac{1}{r}v_r - \frac{1}{r}\partial_\phi v_\phi  &  \partial_r v_\phi - \frac{1}{r}v_\phi + \frac{1}{r}\partial_\phi v_r  & 0\\
0&0&0
\end{array}
\right),
\ee
by identifying $\eta_o (>0)$ with $-\eta_3$. 
{$\eta_o$ also appears as coefficient $-\eta_1^o$
in \citep{Khain2022}. Since the liquid is three-dimensional, there is a third
velocity component $v_z$ related to $v_r$ and $v_\phi$ through the isochoric constraint
\be
\partial_r(rv_r) + \partial_\phi v_\phi + r \partial_z v_z =0.  
\ee
For the sake of clarity, in \rr{sigma0} we have chosen only one of the coefficients that appear in the stress \rr{seta1234}
to characterize our odd viscous liquid.
This is done so that the fluid flow behavior associated with this coefficient becomes uncoupled to 
other types. 
It would be possible to also consider nonzero $\eta_4$ (corresponding
to $\eta^o_2$ in \citep{Khain2022}). We thus revisit the current problem in section \ref{sec: eta4}
by considering the $\eta_4$ viscosity coefficient in \rr{seta1234} both individually and in conjunction
to $\eta_o$. 

In Appendix \ref{sec: oddstress} we have formulated the odd stress \rr{sigma0} 
in terms of the rate-of-strain tensor in two and in three dimensions which follows more closely the 
continuum mechanics literature and we employ this formulation to establish the non-dissipative
nature of the odd stress. }

\section{Three-dimensional waves in an odd viscous liquid}
\subsection{\label{sec: inertial1}Axisymmetric inertial waves}
In this article we tacitly assume the existence of an axis of anisotropy in the $z$-direction, established by 
a secondary mechanism such as a magnetic field or rotation, to which, however, we make no reference.  
Consider an inviscid liquid endowed with odd viscosity as in \rr{sigma0} and an axially symmetric wave propagating along
the axis of the magnetic field. Following \cite[\S14]{Landau1987}, we 
consider cylindrical polar coordinates $r, \phi, z$ (cf. Fig. \ref{cylinder}), the fields are independent of $\phi$, 
we neglect nonlinear terms (assuming small-amplitude motions)
and the time and axial dependence are given by the factor $\exp [ i (kz-\omega t)]$
where the frequency $\omega$ and wave number $k$ along the axis are both real.  
Employing the 
constitutive law \rr{sigma0}, the linearized equations of motion (see Appendix \ref{sec: appendix1}) become
\begin{align} \label{nsr}
-i\omega v_r & = -\frac{1}{\rho}\frac{\partial p'}{\partial r} 
- \nu_o\left[ \frac{1}{r} \frac{\partial}{\partial r} \left( r \frac{\partial v_\phi}{\partial r}\right)   - \frac{v_\phi}{r^2}\right],  \\
-i\omega v_\phi & = 
\nu_o\left[ \frac{1}{r} \frac{\partial}{\partial r} \left( r \frac{\partial v_r}{\partial r}\right)   - \frac{v_r}{r^2}\right],  
\label{nsphi}
\\
-i\omega v_z& = -\frac{i k}{\rho} p', \label{nsz}
\end{align}
where $p'$ is the variable part of the pressure in the wave and $\rho$ is the liquid's constant density. The equation of continuity is
\be \label{incrphi}
\frac{1}{r} \frac{\partial}{\partial r} \left( r v_r\right) + i kv_z= 0. 
\ee
Because of Eq. \rr{nsz} and continuity, $p'/\rho = \omega v_z /k = \frac{i\omega}{k^2} \frac{1}{r} \frac{\partial}{\partial r} \left( r v_r\right)$ and the identity
$\frac{\partial}{\partial r} \left(\frac{1}{r} \frac{\partial}{\partial r} \left( r v_r\right)\right) = 
\frac{1}{r} \frac{\partial}{\partial r} \left( r \frac{\partial v_r}{\partial r}\right)   - \frac{v_r}{r^2}$, 
we obtain
$
\frac{1}{\rho}\frac{\partial p'}{\partial r} = \frac{i\omega}{k^2} \left[   \frac{1}{r} \frac{\partial}{\partial r} \left( r \frac{\partial v_r}{\partial r}\right)   - \frac{v_r}{r^2}       \right]. 
$
Thus, introducing the linear operator
\be \label{L}
\mathcal{L} = \partial_r^2 + \frac{1}{r}\partial _r - \frac{1}{r^2}
\ee
the $r$ and $\phi$ momentum equations become
\begin{align} \label{systemL1}
-i\omega v_r &= - i\frac{\omega}{k^2} \mathcal{L} v_r - \nu_o\mathcal{L} v_\phi, \\
-i\omega v_\phi &= \nu_o\mathcal{L} v_r. \label{systemL2}
\end{align}
Expressing the velocities $v_r$ and $v_\phi$ in terms of Bessel or modified Bessel functions, 
$v_r = AJ_m(\kappa r), v_\phi = BJ_m(\kappa r)$, or $v_r=AI_m(\kappa r), v_\phi =BI_m(\kappa r)$, etc., 
(where $A$ and $B$ are constants and $\kappa$ is an
eigenvalue) we find that $m=1$. With the identity $\mathcal{L}J_1(\kappa r) = - \kappa^2 J_1(\kappa r)$, 
the system \rr{systemL1} and \rr{systemL2} has a solution when the determinant 
$\kappa^4 k^2\nu_o^2 - \omega^2(\kappa^2 + k^2)$
of the coefficients 
of the resulting linear system
\be
Ak^2\omega - \kappa^2(i\nu_o Bk^2 - A\omega) =0, \quad \textrm{and} 
\quad i\nu_o A\kappa^2 +\omega B = 0
\ee
vanishes. 
Consider first the case where the origin is included in the domain. It is not difficult to show that the solution is the Bessel function $J_1(\kappa r)$ for which $\kappa$ satisfies
\be \label{lambda2}
\kappa^2 = \omega \frac{\omega \pm \sqrt{\omega^2 + (2k^2\nu_o)^2}}{2k^2\nu_o^2}.
\ee

Thus, overall we found
\be \label{vrvphi}
v_r = AJ_1(\kappa r)e^{i (kz-\omega t)}, \quad v_\phi = -iA\frac{\nu_o\kappa^2}{\omega} J_1(\kappa r)e^{i (kz-\omega t)}, \quad v_z = iA\frac{\kappa}{k} J_0(\kappa r)e^{i (kz-\omega t)}.
\ee
The motion comprises regions between coaxial cylinders with radius $r_n$ such that
\be
r_n\kappa = \gamma_n,
\ee
and $\gamma_n$ are the zeros of $J_1(x)$. Both $v_r$ and $v_\phi$ vanish at these coaxial cylinders and the fluid does not 
cross them. 
The allowed values of the frequency $\omega$ in Eq. \rr{lambda2}
are not restricted in any way in the infinite medium under consideration (in contrast, in the rotating fluid case where 
$\kappa = k\sqrt{\frac{4\Omega^2}{\omega^2} -1}$, the angular velocity $\Omega$ of the liquid
is required to satisfy the bound $\omega <2\Omega$, for the solution to be finite). In defining Eq. \rr{lambda2} we have tacitly assumed
that the solutions we pursue are finite in the radial direction $r$ and have thus discarded $\kappa$'s
associated with (exponentially increasing/decreasing) modified Bessel function solutions. $\kappa$'s 
in our discussion are always real (this can be justified by choosing large $\omega$ for instance).

%-------------------------------- 
% Matlab figure with odd_axial_waves_streamlines.m in JFM Journal style files folder
%--------------------------------
\begin{figure}
\vspace{-5pt}
\begin{center}
\includegraphics[height=2.5in,width=3.2in]{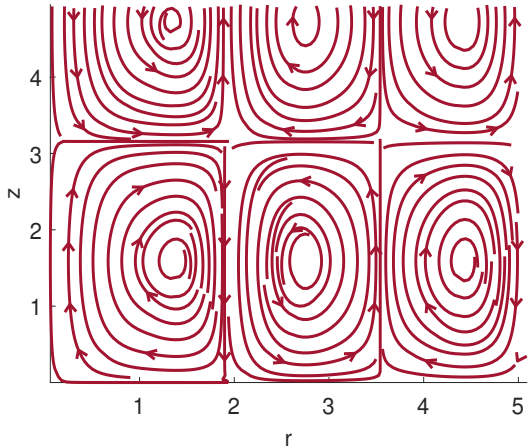}%{odd_axial_psi_zr}
\vspace{-0pt}
\end{center}
\caption{
Instantaneous streamlines in the $r-z$ plane with streamfunction \rr{psi}, representing a simple harmonic wave propagating in the $z-$ direction with
phase velocity $c_p = \omega/k$ (see Eq. \rr{cpcg0}). Vertical lines are cross-sections of cylinders wrapping
around the central axis and were formed from the zeros of the Bessel function $J_1$ where $v_r =0$. 
The radius $b$ of the external cylinder is determined by the condition $\kappa b = \gamma_3$, where
$\gamma_3$ is the third root of the Bessel function $J_1$ in the streamfunction \rr{psi} or equivalently in the
radial velocity field in Eq. \rr{vrvphi}.
\label{psi_zr}  }
\vspace{-0pt}
\end{figure}

Employing the radial and azimuthal momentum equations \rr{nsr} and \rr{nsphi} we can now give 
a more satisfying explanation of the ``elasticity'' of an odd viscous liquid and its tendency to restore
a fluid particle back to its original position. Following \citep[\S 1.1]{Davidson2013book} and \citep[\S 5]{Yih1988}, consider a circular ring of fluid in an odd viscous liquid with zero shear viscosity, wholly 
located on the $x$-$y$ plane. 
{By some perturbation, the ring starts moving outwards with velocity $v_r>0$
and thus expands, so that $\partial_r v_r + \frac{1}{r}v_r + \frac{1}{r}\partial_\phi v_\phi>0$, where the 
last expression is the divergence of the velocity field in two dimensions. Rewrite Eq. \rr{systemL1}-\rr{systemL2} as 
\be \label{systemL}
-i\omega \left(1 + \frac{\kappa^2}{k^2}\right) v_r= F_r, \quad -i\omega v_\phi = F_\phi
\ee
where $F_r \equiv - \nu_o\mathcal{L} v_\phi = \nu_o \kappa^2 v_\phi $ and 
$F_\phi \equiv \nu_o\mathcal{L} v_r = -\nu_o\kappa^2 v_r$. Since $v_r>0$ the second of 
\rr{systemL} implies that there will be an azimuthal force 
$F_\phi = -\kappa^2 \nu_o v_r<0$, an acceleration of the liquid in the $-\hat{\bm{\phi}}$
direction and a commensurate negative velocity $v_\phi$, where we employed the eigenvalue $-\kappa^2$
of the linear operator $\mathcal{L}$ in \rr{L}. This velocity 
will give rise to a radial force $F_r = - \kappa^2 \nu_o |v_\phi|$ in the first of \rr{systemL}. 
This force endows
the ring with an acceleration that points towards the origin, that is, towards the original location of the 
fluid ring, so it tries to reverse its expansion (the pressure contributes the $-i\omega\frac{\kappa^2}{k^2}$
in \rr{systemL}). As the ring passes through its original position due to inertia 
and contracts $\partial_r v_r + \frac{1}{r}v_r + \frac{1}{r}\partial_\phi v_\phi<0$,} 
there will be a new azimuthal velocity component
with opposite sign to the above one, that will lead to an eventual expansion towards equilibrium.

\subsection{Axial inertial waves interior to a cylinder}
We consider the liquid confined within a solid cylindrical surface located, say, at $r=a$, that would be realistic in a laboratory setting. 
This boundary will be a streamline located at an integral number of cells in the radial direction. If by $\gamma_n$ we denote
the $n$-th zero of the Bessel function $J_1$, Eq. \rr{lambda2} with the condition 
$\kappa a= \gamma_n$ lead to the constraint
\be \label{7.6.8}
a\left(\omega \frac{\omega + \sqrt{\omega^2 + (2k^2\nu_o)^2}}{2k^2\nu_o^2}\right)^{1/2} = \gamma_n,
\ee
and $n$ denotes the number of cells in the radial direction (cf. Fig. \ref{psi_zr}). 
From Eq. \rr{7.6.8} we derive the dispersion relation
\be \label{omega}
\omega  = \frac{\nu_o k\gamma_n^2}{a \sqrt{k^2a^2 + \gamma_n^2}} ,
\ee
where $n$ denotes the number of cells in the radial direction 
{(for clarity we have suppressed the symbol $\pm$ in
\rr{omega} and only consider the positive sign).} 
It is clear that in the limit $ka\gg \gamma_n$, the frequency in Eq. \rr{omega} becomes $\omega \sim \frac{\nu_o}{a^2}$ which recovers
the qualitative frequency \rr{omega0} we obtained
at the Introduction of this article. 

In Fig. \ref{psi_zr} we plot the streamlines interior to a cylinder of radius $b$ with the instantaneous streamfunction
\be \label{psi}
\psi(r,z) = \frac{\kappa}{3.83} rJ_1(\kappa r) \sin (kz), \quad v_z = \frac{1}{r}\frac{\partial \psi}{\partial r}, \quad 
v_r = -\frac{1}{r}\frac{\partial \psi}{\partial z},
\ee
for $k=1$ and $\kappa = 2$ for comparison with Fig. 7.6.4 of \cite[p.561]{Batchelor1967} 
{who
is employing the same form for the streamfunction with the first zero $3.83$ of the Bessel function
to modulate the amplitude in the denominator of \rr{psi}. }
The horizontal lines are locations where $v_r=0$ (zeros of the Bessel function).

There is important information to be surmised from the phase and group velocities
\be \label{cpcg0}
c_p = \frac{\nu_o \gamma_n^2}{a \sqrt{k^2a^2 + \gamma_n^2}}, \quad \textrm{and} \quad 
c_g =\frac{\nu_o \gamma_n^4}{a \left(k^2a^2 + \gamma_n^2\right)^{3/2}},
\ee
{we derive from Eq. \rr{omega} (we have suppressed the symbol $\pm$ and employed only the positive 
sign in \rr{omega}).  } 
Since
\be \label{cpcg}
c_p = c_g + \frac{a\nu_o k^2 \gamma_n^2}{ \left(k^2a^2 + \gamma_n^2\right)^{3/2}} > c_g,
\ee
the energy of a disturbance caused by a slowly-moving body along the axis of the cylinder with a 
velocity equal to $c_p$, cannot advance upstream relative to the body. Waves will be formed in the downstream direction. {We reach the analogous conclusion if we employ the negative sign
in \rr{omega}.}
This situation is thus similar to the rotating liquid case where the energy cannot propagate upstream and thus
waves are formed only downstream, as described in many experiments, e.g. those of \citep{Long1953}. 
We verify these claims in section \ref{sec: inertial} by combining numerical simulations of the Navier-Stokes
equations with the theoretical predictions of the present section. 
(Note how the expression for $\omega$ in Eq. \rr{omega} contrasts with the inviscid liquid rotating at an angular velocity $\Omega$, where $\omega = 2\Omega k/\sqrt{k^2 + \left(\frac{\gamma_n}{a}\right)^2}$).

\subsection{\label{sec: inertial2}Allowed wave-numbers}
When the number of cells in the radial direction is $n$, 
from \rr{7.6.8} allowed wavenumbers  supporting propagation with phase velocity $c_p=\omega/k$ 
and for which the boundary at $r=a$ is a streamline, satisfy
\be \label{bk}
ak = \gamma_n\left( \left(\frac{\nu_o\gamma_n}{ac_p}\right)^2 - 1\right)^{1/2}.
\ee
Wave propagation is thus possible when
\be \label{bcp1}
\frac{ac_p}{\nu_o} < \gamma_n
\ee
where $n$ denotes the number of cells in the radial direction. Thus, although Eq. \rr{lambda2} does not
introduce a restriction on frequencies for the propagation of waves, Eq. \rr{bk} does: 
Defining a Maxworthy number $\mathcal{M}_a$ based on the phase velocity $c_p$ and cylinder radius $a$
(cf. Eq. \rr{MaxworthyM} for the definition of the dimensionless number $\mathcal{M}$)
\be
\mathcal{M}_a = \frac{\nu_o}{ac_p}, 
\ee
inertial motions with $n$ cylinders ($n$-th zero of $J_1$) are possible only when
\be
\mathcal{M}_a > \frac{1}{\gamma_n}. 
\ee
%Thus, if $\frac{1}{\gamma_{n}}< \mathcal{M}_b < \frac{1}{\gamma_{n-1}}$, 
%a motion with $n$ cylinders (or more) is possible (the index $n$ corresponds to $n$-th zero of $J_1$). 

\begin{figure}
%   \centering
    \begin{subfigure}[t]{0.55\textwidth}
%        \centering
        \includegraphics[width=\linewidth]{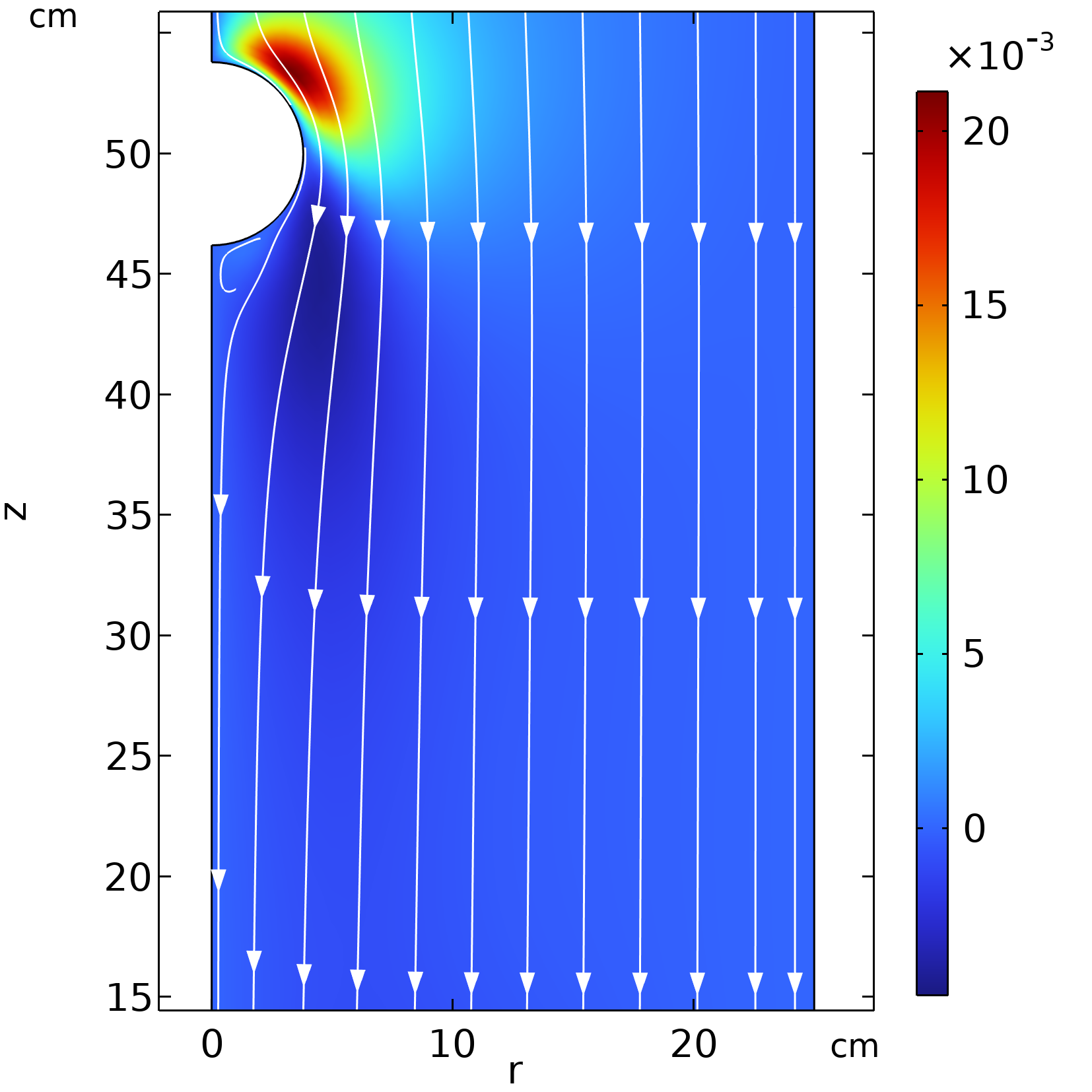} 
%        \caption{} \label{chiraltorque1}
    \end{subfigure}
%    \hfill
    \begin{subfigure}[t]{0.55\textwidth}
%       \centering
        \includegraphics[width=\linewidth]{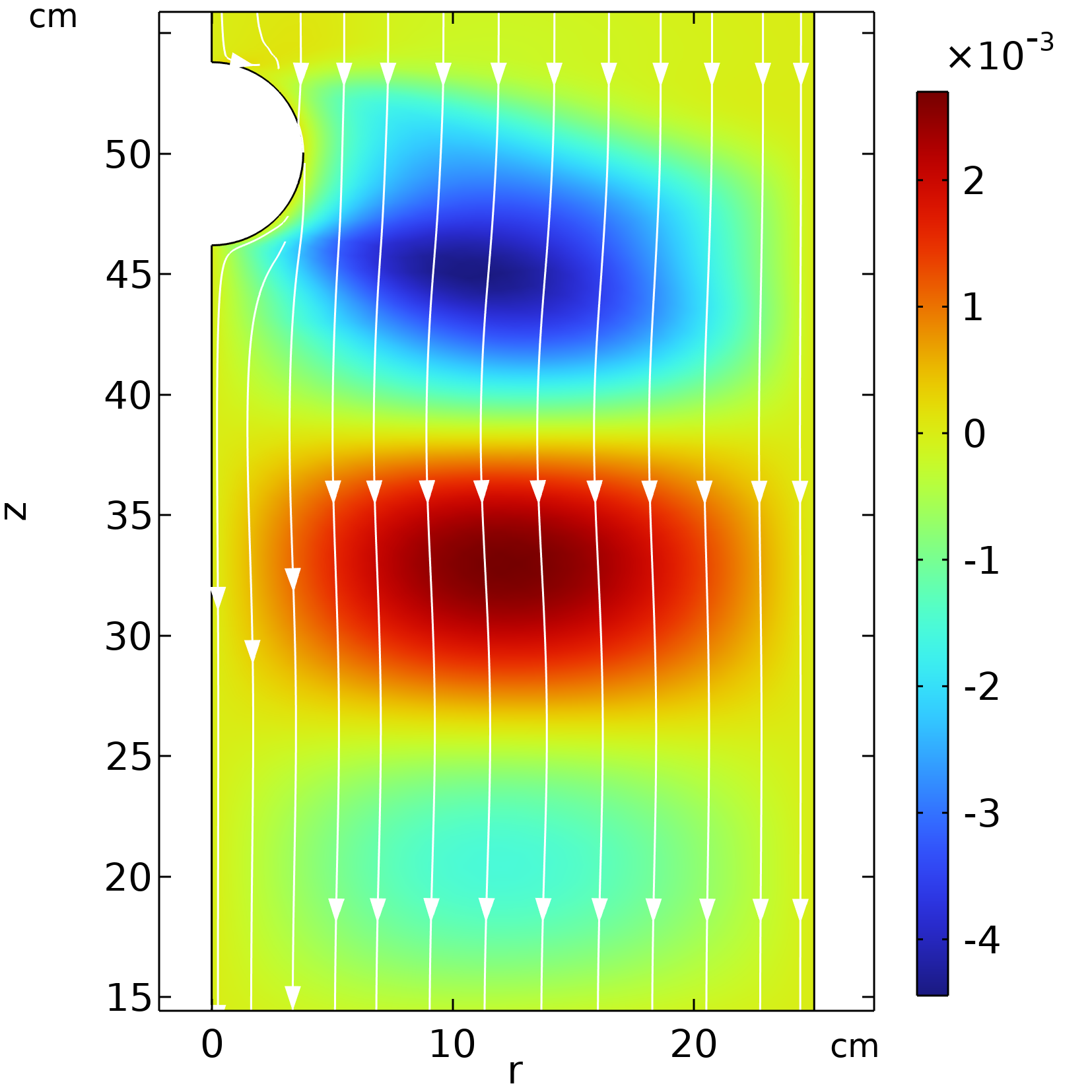} 
%        \caption{Spatial distribution of the observables $v_p, \sigma^{\textrm{ch}}$ and $v^{\textrm{ch}}$.} \label{vpschvchN}
    \end{subfigure}
     %  \vspace{1cm}
%    \begin{subfigure}[t]{\textwidth}
%    \centering
%        \includegraphics[width=\linewidth]{example-image-c.pdf} 
%        \caption{Price regulation} \label{fig:timing3}
%    \end{subfigure}
    \caption{\textbf{Right:} Waves generated by a small ($3.8$ cm) slowly-moving sphere (located at the center of the cylinder - 
upper left of the figure) in an odd viscous liquid contained in a cylinder of radius $25$ cm. The color bar denotes 
the strength of the radial liquid velocity $v_r$. Its direction changes sign as we move down and 
away from the body and it is thus responsible for the distortion of the streamlines (in white). 
From this plot we can visually determine the wavelength to approximately be $25$ cm long. This 
agrees rather well with the
theoretical estimate of $24.476$ cm obtained from Eq. \rr{wavelength}. {\textbf{Left:} In the absence 
of odd viscosity no radial disturbance is visible as we move down and 
away from the body. Colorbar denotes radial velocity}.  
The figure was produced with the finite-element package comsol by solving the full Navier-Stokes equations
including inertial terms in a three-dimensional axisymmetric domain. 
\label{oscillations1}} 
\end{figure}

\subsection{\label{sec: inertial}Numerical determination of odd viscous inertial oscillations inside a cylinder of radius $a$}
By the inequality \rr{cpcg} we argued that, based on the inertial waves construction of sections 
\ref{sec: inertial1} - \ref{sec: inertial2},  
the energy of a disturbance caused by a slowly-moving body along the axis of the cylinder with a 
velocity equal to $c_p$, cannot advance upstream relative to the body. Waves will be formed in the downstream direction. 

To verify this claim we perform numerical simulations of the full Navier-Stokes equations of a slowly-moving
body with velocity $U=0.05$ cm/sec in a cylinder
of base radius $a = 25$ cm filled with an odd viscous liquid of dynamic coefficient $\eta_o = 0.7$ g/(cm sec), shear
viscosity $\eta = 0.01$ g/(cm sec) and density $\rho = 1.1$ g/$\textrm{cm}^3$.
Figure \ref{oscillations1} displays the streamlines in the $r-z$ plane of the liquid downstream the moving
body which show wave-like behavior. To determine the wavelength, the color-bar shows 
the strength of the radial liquid velocity $v_r$ whose direction changes sign as we move down.
From the plot we can visually determine that the wavelength $\lambda$ is approximately $25$ cm long. 
We compare this estimate
to the theoretical prediction of sections \ref{sec: inertial1} - \ref{sec: inertial2}: From Eq. \rr{bk} we
find that 
\be \label{wavelength}
\lambda = \frac{2\pi}{k} = \frac{2\pi a}{\gamma_1 \sqrt{\left(\frac{\nu_o \gamma_1}{aU}\right)^2 - 1}}
=24.4764 \textrm{ cm}
\ee
where $\gamma_1 = 3.8317$ is the first root of the Bessel function $J_1$. 
Figure \ref{oscillations1} was produced with the finite-element package comsol by solving the full 
Navier-Stokes equations
in a three-dimensional axisymmetric domain. The flow Reynolds number, based on cylinder radius $a=25$ cm is 125.

This situation is thus analogous to the rotating liquid case where the energy cannot propagate upstream and thus
waves are formed only downstream, as described by the experiments of \citet{Long1953}, cf. 
\citep{Batchelor1967} pp.564-6 and plate 24.

\subsection{\label{sec: exterior}Axial inertial waves exterior to a cylinder}
\begin{figure}
\vspace{-5pt}
\begin{center}
\includegraphics[height=2.4in,width=3in]{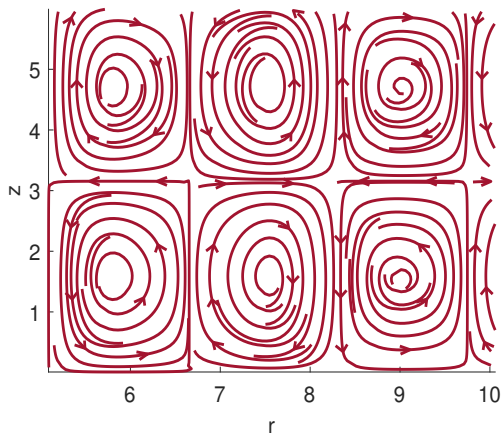}%{odd_axial_psi_zr2}
\vspace{-0pt}
\end{center}
\caption{
Instantaneous streamlines exterior to a cylinder of radius $r=\frac{\alpha_3}{\kappa}$ where $\alpha_3$ is the third zero of $Y_1$. This is simple harmonic wave propagating in the $z-$ direction with
phase velocity $\omega/k$. Vertical lines correspond to zeros of the Bessel function of second kind $Y_1$ where $v_r =0$. 
\label{psi_zr2}  }
\vspace{-0pt}
\end{figure}

The above discusion can also be employed to establish wave propagation when
the odd viscous liquid occupies the region $r>a$, exterior to a solid cylinder located at $r=a$. 
Now, the solution is of the form of a Bessel function of the second kind in the radial coordinate
\be
v_r = AY_1(\kappa r)e^{i (kz-\omega t)}, \quad v_\phi = -iA\frac{\nu_o\kappa^2}{\omega} Y_1(\kappa r)e^{i (kz-\omega t)}, \quad v_z = iA\frac{\kappa}{k} Y_0(\kappa r)e^{i (kz-\omega t)}
\ee
and the results of the previous sections hold with the replacement
\be
J_1 \rightarrow Y_1, \quad \gamma_n \rightarrow \alpha_n, 
\ee
where $\alpha_n$ is the $n$th zero of $Y_1(x)$.
In Fig. \ref{psi_zr2} we plot the streamlines exterior to a cylinder of radius $b$ with the instantaneous streamfunction
\be \label{psi2}
\psi(r,z) = \frac{\kappa}{3.83} rY_1(\kappa r) \sin (kz), \quad v_z = \frac{1}{r}\frac{\partial \psi}{\partial r}, \quad 
v_r = -\frac{1}{r}\frac{\partial \psi}{\partial z}
\ee
for $k=1$ and $\kappa = 2$ for comparison with Fig. 7.6.4 of \cite[p.561]{Batchelor1967}.

\subsection{\label{sec: plane}Plane-polarized waves induced by odd viscosity}
The axisymmetric inertial waves we discussed earlier propagate along the $z-$ axis (the axis of anisotropy) and are three-dimensional in the sense that the wave amplitude varies along both the propagation direction
and normal to it.
In this section we will consider different types of inertial waves that propagate
along an arbitrary direction $\mathcal{k}$ and are polarized in the plane perpendicular to the 
propagation axis. This section follows the notation of \cite[\S 14]{Landau1987}.
The odd-viscous Navier-Stokes equations
\be \label{DvDt}
\frac{D\mathbf{v}}{Dt} = -\frac{1}{\rho}\nabla p +\nu_o \nabla_2^2\hat{ \mathbf{z}} \times \mathbf{v},
\ee
where $\nabla^2_2 = \partial_x^2 + \partial_y^2$ and $D/Dt$ is the convective derivative, become after
taking the curl of both sides 
\be \label{dtcurlvnl}
\frac{D}{Dt}\textrm{curl} \mathbf{v} =\textrm{curl} \mathbf{v} \cdot \nabla \mathbf{v} - \nu_o \nabla_2^2 \frac{\partial \mathbf{v}}{\partial z}. 
\ee
Linearizing, 
\be \label{dtcurlv}
\partial_t\textrm{curl} \mathbf{v} = - \nu_o \nabla_2^2 \frac{\partial \mathbf{v}}{\partial z},
\ee
we seek plane-wave solutions of the form
\be\label{vA}
\mathbf{v} = \mathbf{A} e^{i(\mathbf{k}\cdot \mathbf{r} - \omega t)}, 
\ee
where $\mathbf{A}$ is normal to $\mathbf{k}$ from the incompressibility condition.

\begin{figure} 
\begin{center}
\includegraphics[height=3in,width=3in]{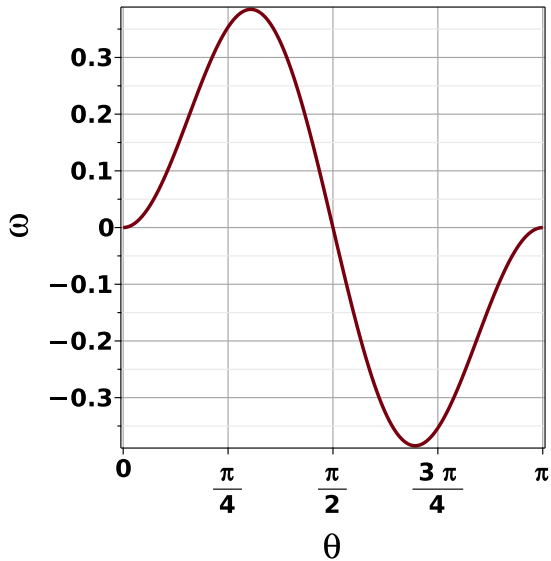}
\end{center}
\vspace{-5pt}
\caption{Section \ref{sec: plane} plane-polarized wave dispersion $\omega$ (Eq. \rr{disp1}) versus angle $\theta$ between the wavevector $\mathbf{k}$
and the $z$- (anisotropy) axis (setting $k=\nu_o = 1$).  Of interest is the low frequency range at
$\theta \sim \frac{\pi}{2}$ where group velocity is maximum. This curve structure should be contrasted to the corresponding
relation $\omega = 2\Omega \cos \theta$ of an inviscid fluid rotating with angular velocity $\Omega$.
\label{omega1} }
\end{figure}

Substituting the plane-wave solution into \rr{dtcurlv} we obtain
\be \label{kcrossv1}
\omega \mathbf{k} \times \mathbf{v} = i \nu_o (k_x^2 + k_y^2) k_z\mathbf{v}.
\ee
Taking the cross product of both sides of Eq. \rr{kcrossv1} with $\mathbf{k}$ we obtain
\be \label{kcrossv2}
-\omega k^2 \mathbf{v} = i \nu_o (k_x^2 + k_y^2) k_z\mathbf{k} \times \mathbf{v}.
\ee
System \rr{kcrossv1}-\rr{kcrossv2} has a solution when the determinant of the coefficients vanishes. 
Solving for $\omega$ we obtain 
\be \label{disp1}
\omega = \pm \frac{\nu_o (k_x^2 + k_y^2) k_z}{k}, \quad \textrm{or} \quad 
\omega = \pm \nu_o k^2 \cos\theta \sin^2\theta, 
\ee
where $k = \sqrt{k_x^2 + k_y^2 + k_z^2}$
and the latter equation implies that $\theta$ is the angle between the $\mathbf{k}$ and the anisotropy axis. 
In Fig. \ref{omega1} we plot the dispersion $\omega$ (Eq. \rr{disp1}) versus angle $\theta$ between the wavevector $\mathbf{k}$ and the $z$- (anisotropy) axis. 
{
It differs qualitatively from the corresponding
relation 
\be \label{rotom}
\omega = 2\Omega \cos \theta 
\ee
of a (non-odd viscous) inviscid fluid rotating with angular velocity $\Omega$,
cf. \citep{Greenspan1968}. 
Comparing Eq. \rr{disp1} with \rr{rotom}, when the coefficients $\nu_o$ and $\Omega$ are kept constant, it becomes evident that the dispersion relation \rr{disp1} obtained due to the specific form of the 
constitutive law \rr{sigma0} we adopted for an odd viscous liquid becomes prominent for 
large in-plane wave numbers and small corresponding wavelengths. 
}

Following \cite{Landau1987}, we introduce the unit vector $\hat{\mathbf{k}} = \frac{\mathbf{k}}{k}$
in the direction of the wave-vector and the complex amplitude $\mathbf{A} = \mathbf{a} +i \mathbf{b}$ 
where $\mathbf{a}$ and $\mathbf{b}$ are real vectors. 
Considering Eq. \rr{kcrossv1} and the dispersion relation \rr{disp1} we obtain
$\hat{\mathbf{k}} \times \mathbf{b} = \mathbf{a}$, that is, the two vectors $\mathbf{a}$ and $\mathbf{b}$ are perpendicular to each other, are of the 
same magnitude and lie in the plane whose normal is $\mathbf{k}$. Thus, the velocity field is 
circularly polarized in the plane defined by $\mathbf{a}$ and $\mathbf{b}$ and is of the form
\be \label{vab}
\mathbf{v} = \mathbf{a} \cos (\mathbf{k}\cdot \mathbf{r} - \omega t) - \mathbf{b} \sin (\mathbf{k}\cdot \mathbf{r} - \omega t), \quad \mathbf{a}\perp\mathbf{b}.
\ee
Employing the negative sign of the dispersion relation \rr{disp1}, the above analysis leads to the same
velocity field \rr{vab} but with the sense of the vectors  $\mathbf{a}$ and $\mathbf{b}$ reversed:
$\hat{\mathbf{k}} \times \mathbf{b} = -\mathbf{a}$. This will become important in section 
\ref{sec: helicity} where the helicity associated with wave propagation will be determined.

\begin{figure} 
\begin{center}
\includegraphics[height=2.2in,width=5in]{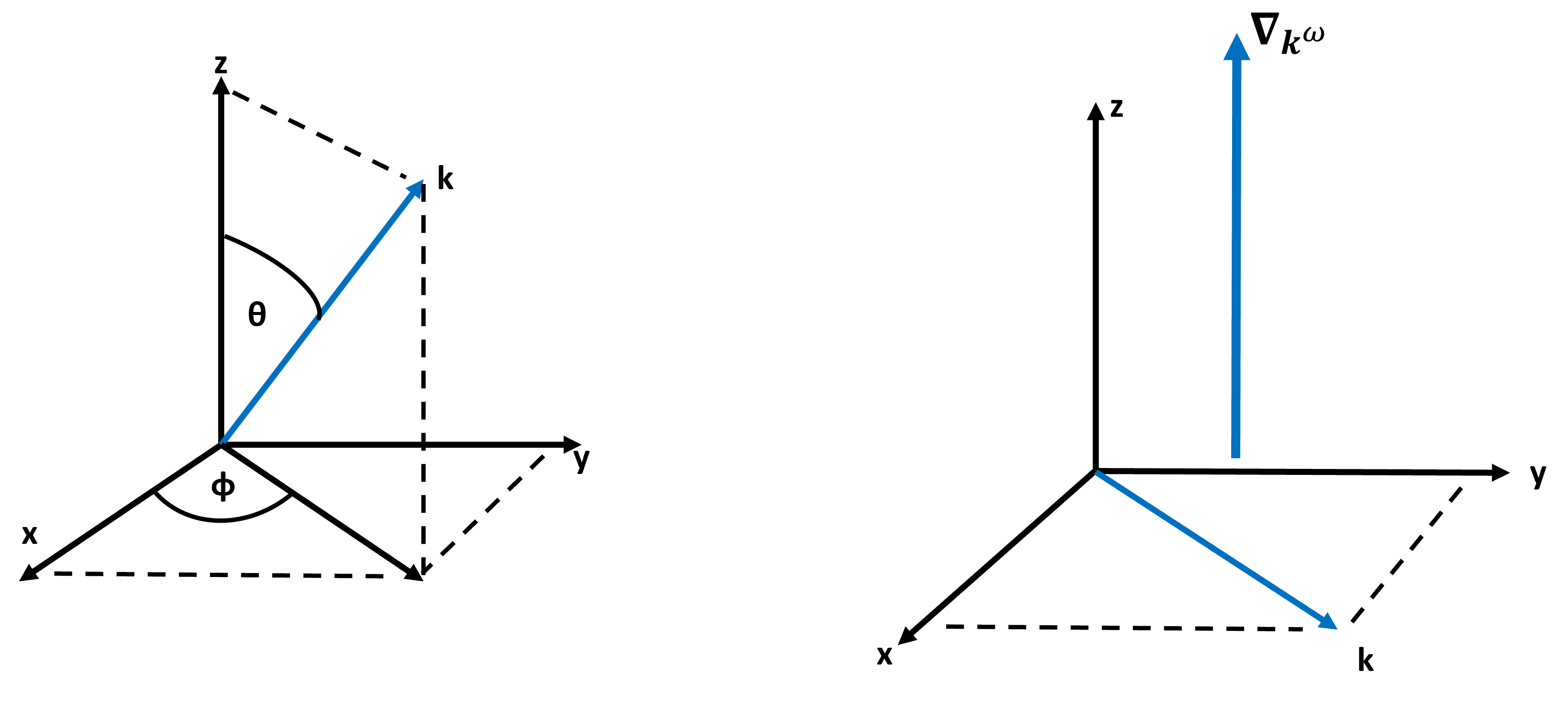}
\end{center}
\vspace{-5pt}
\caption{\textbf{Left:} coordinate system employed in plane-polarized waves showing the definition
of angles for the propagation wavevector $\mathbf{k}$. \textbf{Right:} When the propagation direction
is normal to the axis $\hat{\mathbf{z}}$, the group velocity $\mathbf{c}_g = \nabla_\mathbf{k}\omega$
in \rr{polgroup2} becomes co-axial to the axis $\hat{\mathbf{z}}$ and 
acquires its maximum value.   
\label{coordinate1} }
\end{figure}

\begin{figure} 
\begin{center}
\includegraphics[height=3in,width=3in]{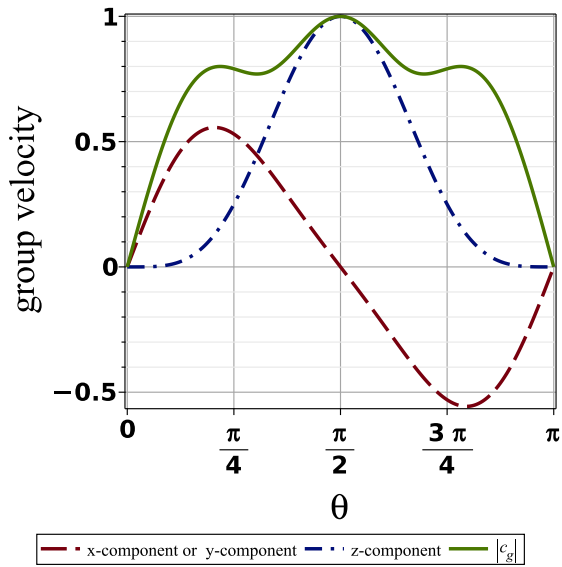}
\end{center}
\vspace{-5pt}
\caption{Section \ref{sec: plane} plane-polarized wave group velocity components and magnitude Eq. \rr{polgroup1} and \rr{polgroup2} versus angle $\theta$ between the wavevector $\mathbf{k}$
and the $z$- (anisotropy) axis, setting $k=\nu_o = 1$ (and $ \phi =\pi/4$, for simplicity). 
Of interest is the low frequency range at
$\theta \sim \frac{\pi}{2}$ where group velocity is maximum. 
The modulus of 
the group velocity should be contrasted to $\left|\frac{\partial \omega}{\partial \mathbf{k}} \right|= (2\Omega/k) \sin\theta$ in the case of an inviscid rigidly rotating liquid with angular velocity $\Omega$. 
\label{cg1} }
\end{figure}

It is of interest to calculate the direction of propagation of energy. We obtain
\be
\frac{\partial \omega}{\partial k_x} = 
\nu_o\frac{k_xk_z(k^2 + k_z^2)}{k^3}, \quad 
\frac{\partial \omega}{\partial k_y} = 
\nu_o\frac{k_yk_z(k^2 + k_z^2)}{k^3}, \quad
\frac{\partial \omega}{\partial k_z} = 
\nu_o\frac{(k_x^2 + k_y^2)^2}{k^3}
\ee
or, taking the $z$ axis to be the axis of anisotropy we obtain
\be \label{polgroup1}
\left(\frac{\partial \omega}{\partial k_x}, \frac{\partial \omega}{\partial k_y} \right)
= k\nu_o \sin \theta  \cos \theta (1 + \cos^2 \theta)( \cos \phi, \sin \phi),
\quad 
\frac{\partial \omega}{\partial k_z} = k\nu_o \sin^4 \theta. 
\ee
%\be \label{polgroup1}
%\frac{\partial \omega}{\partial k_x} = k\nu_o \sin \theta  \cos \theta (1 + \cos^2 \theta)\cos \phi, 
%\quad 
%\frac{\partial \omega}{\partial k_y} = k\nu_o \sin \theta  \cos \theta (1 + \cos^2 \theta)\sin \phi, 
%\quad
%\frac{\partial \omega}{\partial k_z} = k\nu_o \sin^4 \theta. 
%\ee
{
The group velocity $\mathbf{c}_g = \frac{\partial \omega}{\partial \mathbf{k}}$ in vector form 
can be written as 
%In contrast to the case of waves in a rotating fluid where the energy propagates perpendicularly to the 
%wave-vector, 
%here the energy propagation direction has a component
%along the $\hat{\mathbf{z}}$ axis:
\be
\frac{\partial \omega}{\partial \mathbf{k}} = \nu_ok 
\left\{   \hat{\mathbf{k}}(\hat{\mathbf{z}}\cdot \hat{\mathbf{k}}) \left[ 1+ (\hat{\mathbf{z}}\cdot \hat{\mathbf{k}})^2\right] +  \hat{\mathbf{z}}\left[1- 3(\hat{\mathbf{z}}\cdot \hat{\mathbf{k}})^2 \right]
\right\},
\ee
which can be compared with its rigidly-rotating (non-odd viscous) counterpart 
$\frac{\partial \omega}{\partial \mathbf{k}} = \frac{2\Omega}{k}\left[\hat{\mathbf{z}} -  \hat{\mathbf{k}}(\hat{\mathbf{z}}\cdot \hat{\mathbf{k}}) \right]$ where the group velocity is perpendicular to the 
phase velocity $\mathbf{c}_p = \frac{\omega}{k} \hat{\mathbf{k}}$ (see Table \ref{table:table1}).  
Here the group velocity is not perpendicular to the phase velocity. A calculation gives 
$\mathbf{c}_g \cdot \mathbf{c}_p = 2|\mathbf{c}_p|^2 = 2\nu_o^2k^2\cos^2\theta \sin^4\theta$. Thus, in contrast to the case of inertial waves in a rotating fluid where the energy propagates perpendicularly to the 
wave-vector, 
here the energy propagation direction has a component
along the $\hat{\mathbf{k}}$ axis. 
}
%\be
%\frac{\partial \omega}{\partial \mathbf{k}} = \nu_o\frac{k_x^2 + k_y^2}{k} 
%\left[ \hat{\mathbf{z}} - \hat{\mathbf{k}} (\hat{\mathbf{z}}\cdot \hat{\mathbf{k}}) \right] + 
%2 \nu_o\frac{k_z^3}{k^2} 
%\left[ \hat{\mathbf{k}} - \hat{\mathbf{z}} (\hat{\mathbf{z}}\cdot \hat{\mathbf{k}}) \right]
%\ee
The modulus of the group velocity is 
\be \label{polgroup2}
\left|\frac{\partial \omega}{\partial \mathbf{k}} \right|= k\nu_o \sin\theta 
\sqrt{5 \cos^4\theta -  2 \cos^2 \theta + 1}. 
\ee
%and along the anisotropy axis
%\be
%\frac{\partial \omega}{\partial \mathbf{k}} \cdot \hat{\mathbf{z}} = \nu_o\frac{k_x^2 + k_y^2}{k} \sin^2\theta
%\ee
In Fig. \ref{cg1} we plot the group velocity components Eq. \rr{polgroup1} and its magnitude \rr{polgroup2} versus angle $\theta$ between the wavevector $\mathbf{k}$
and the $z$- (anisotropy) axis (setting $k=\nu_o = 1$).

In addition to the information of Table \ref{table:table1}, the frequency is maximum 
at $\theta=\cos^{-1}\frac{\sqrt{3}}{3} \sim 0.3\pi$, it acquires the value $\omega = \pm\nu_ok^2\frac{2\sqrt{3}}{9}$. The group velocity becomes $\frac{4k\nu_o}{9}(\sqrt{2} \cos \phi, \sqrt{2} \sin\phi, 1)$ and its 
modulus is $c_g = 4k\nu_o \sqrt{3}/9$.  
In addition the group velocity has a local maximum at $\theta=\cos^{-1}\frac{\sqrt{15}}{5} \sim 0.22\pi$. 
The frequency is $\omega = \pm\nu_ok^2\frac{2\sqrt{15}}{25}$ and the group velocity becomes $\frac{4k\nu_o}{25}(2\sqrt{6} \cos \phi, 2\sqrt{6} \sin\phi, 1)$ and its 
modulus is $c_g = 4k\nu_o / 5$. 
Comparison can be made of Eq. \rr{polgroup2} with plane-polarized waves rigidly rotating \cite[\S 14]{Landau1987} 
% Inclusion of shear viscosity gives rise to damping \cite[p.83]{Chandrasekhar1961}. 
where $\left|\frac{\partial \omega}{\partial \mathbf{k}} \right|= (2\Omega/k) \sin\theta$.

\section{\label{sec: helicity}Conservation of helicity of inertial waves in an odd viscous liquid}
Helicity $\mathcal{H}$, 
\be \label{helicity}
\mathcal{H}  = \int_V \mathbf{v} \cdot \textrm{curl}\mathbf{v} dV= \textrm{constant}, 
\ee
was shown by \citet{Moffatt1969} to be 
an invariant of inviscid fluid motion
when $\hat{\mathbf{n}} \cdot \textrm{curl}\mathbf{v}$ vanishes on any closed orientable surface moving with the liquid. Here, we show that, analogously to the rigidly-rotating case, the vorticity of an odd
viscous liquid
in proportional to the velocity field. For plane-polarized inertial waves this implies that helicity is conserved. 
Since a number of effects appearing in the literature, such as the emission of inertial waves in a turbulent flow \citep[Fig. 3.3(b)]{Davidson2013book} are related to helicity and its sign, we include some discussion below on the presence of helicity
in odd viscosity-induced inertial waves.

\subsection{\label{sec: helicity2}Conservation of helicity in plane-polarized waves of an odd viscous liquid}
From the plane-polarized velocity field \rr{vab} we obtain
$\textrm{curl}\mathbf{v} =-\mathbf{k}\times \mathbf{b} \cos (\mathbf{k}\cdot \mathbf{r} - \omega t) -\mathbf{k}\times \mathbf{a} \sin (\mathbf{k}\cdot \mathbf{r} - \omega t)$. The relation $\hat{\mathbf{k}} \times \mathbf{b} = \pm \mathbf{a}$ (the $\pm$ symbol corresponds to the sign of the dispersion relation \rr{disp1}) leads to $\textrm{curl}\mathbf{v} = \mp k\mathbf{v}$ and thus 
\be \label{helicity2}
\mathbf{v} \cdot \textrm{curl}\mathbf{v} = \mp k|\mathbf{v}|^2 \quad \textrm{and} \quad \mathcal{H}  = \int_V \mathbf{v} \cdot \textrm{curl}\mathbf{v}dV = \mp k|\mathbf{v}|^2V,
\ee
where $|\mathbf{v}|^2=|\mathbf{a}|^2 + |\mathbf{b}|^2$ is the constant magnitude of the 
velocity in Eq. \rr{vab}, $V$ is the volume of the region under consideration and 
$k = \sqrt{k_x^2 + k_y^2 + k_z^2}$.
The negative sign in \rr{helicity2} (corresponding to the positive sign in the dispersion relation \rr{disp1}) is associated with particle paths following left-handed helices and
the positive sign with right-handed helices. Energy propagates along a cone whose normal is the 
vector $\hat{\mathbf{k}}$, (cf. \citep{Davidson2013book} for the case of rigidly rotating liquid). 

\begin{table}
  \begin{center}
\def~{\hphantom{0}}
  \begin{tabular}{lcc}
 &odd viscous liquid & rigidly-rotating liquid\\
&& \\
low $\omega$: & $\mathbf{k}\parallel \bm{\Omega} $; $c_g=0$ and $\mathbf{k} \perp \bm{\Omega} $;
$\mathbf{c}_g=\pm \nu_o k \hat{\mathbf{z}}$ (max)
&$\mathbf{k}\perp \bm{\Omega}$; $\mathbf{c}_g = \frac{\pm 2\Omega\hat{\mathbf{z}}}{k}$ (max) \\
&& \\
high $\omega$: &$\theta=\cos^{-1}\frac{\sqrt{3}}{3} \sim 0.3\pi$&$\mathbf{k}\parallel \bm{\Omega}$; $\mathbf{c}_g =0$\\ 
$\mathbf{c}_p\cdot \mathbf{c}_g $& $2|\mathbf{c}_p|^2$& 0\\
$\mathbf{v} \cdot \textrm{curl}\mathbf{v}$ &$\mp k|\mathbf{v}|^2$&$\mp k|\mathbf{v}|^2$
\end{tabular}
\caption{Summary of odd viscous plane-polarized inertial waves behavior at low and high frequencies \rr{disp1}
and comparison with their rigidly-rotating counterparts. $\mathbf{c}_g$ is the group velocity $\frac{\partial \omega}{\partial \mathbf{k}}$ and $\mathbf{c}_p = \frac{\omega}{k} \hat{\mathbf{k}}$ is the phase velocity. The group velocity is maximum when the angle $\theta = \pi/2$, propagation takes place 
perpendicular to the anisotropy axis ($\hat{\mathbf{z}}$), the group velocity acquires its maximum
propagating along the anisotropy axis and helicity becomes segregated, cf. section \ref{sec: helicity}. 
Helicity density $\mathbf{v} \cdot \textrm{curl}\mathbf{v}$ has the same functional form in both odd viscous and
rigidly-rotating liquids.  
Although in a rigidly rotating liquid the group velocity is always perpendicular to the phase velocity, 
in an odd viscous liquid there is dependence on the angle $\theta$ (see the third line of the Table). 
 $\theta$ denotes the angle between $\bm{\Omega} = \Omega \hat{\mathbf{z}}$ and
the propagation direction $\mathbf{k}$, cf. Fig. \ref{coordinate1}. }
  \label{table:table1}
  \end{center}
\end{table}

Inertial-like waves give rise to maximal helicity. This can be seen by substituting Eq. \rr{helicity2} into the 
the Cauchy-Schwartz inequality
$\mathcal{H}^2\leq \mathcal{E}\mathcal{W}$ expressed in terms of the helicity \rr{helicity} and the energy and enstrophy integrals
\citep{Moffatt1969}
\be
\mathcal{E} = \int_V \mathbf{v}^2 dV , \quad \mathcal{W} = \int_V \textrm{curl}\mathbf{v}^2 dV. 
\ee

As shown by \citet{Moffatt1970} for the case of rigidly-rotating liquids, inertial waves exhibit a loss of reflection symmetry when the energy propagates parallel to the rotation axis (and the phase velocity is perpendicular
to this axis). \citet{Davidson2013book} associates each direction of propagation of energy with one of the
signs of helicity in \rr{helicity2}. Negative sign of helicity for energy propagating in the $+\hat{\mathbf{z}}$
direction and positive sign of helicity for energy propagating in the $-\hat{\mathbf{z}}$. 
This is called the ``segregation of helicity'' and has found applications in problems of magnetohydrodynamics
\citep{Davidson2013book, Davidson2018}. 
The waves that 
correspond to this type of behavior have low frequencies (the frequency $\omega$ in Eq. \rr{disp1} is 
nearly zero). The consequence of this behavior in an odd viscous liquid can more easily be seen by going back to the original equation
of motion \rr{DvDt}. Linearizing, and taking the limit $\omega \rightarrow 0$ amounts to dropping
the time derivative. Then, the equation of motion becomes \rr{TP1} of the foregoing section, that makes the 
dynamics effectively two-dimensional (perpendicular to the anisotropy axis), leads to the Taylor-Proudman 
theorem and gives rise to Taylor columns.

\subsection{\label{sec: helicity3}Helicity in axisymmetric inertial waves of an odd viscous liquid}
It turns out that vorticity is also parallel to the velocity field for the inertial waves of section \ref{sec: inertial1}. 
This can be shown directly by taking the curl of the velocity field \rr{vrvphi} and solving 
in expression \rr{lambda2} for the frequency, 
$\omega = \frac{\nu_o k\kappa^2}{\sqrt{k^2 + \kappa^2}}$ (in contrast to the previous section, $k$ here denotes the wave number $k_z$ along the axis, cf. section \ref{sec: inertial1}). Alternatively, setting $A = ae^{i\theta}$
for real amplitude $a$ and phase $\theta$ we obtain
\be \label{vreal}
v_r = aJ_1(\kappa r)\cos(kz-\omega t+\theta), \quad v_\phi = \frac{\nu_o \kappa^2}{\omega}aJ_1(\kappa r)\sin(kz-\omega t+\theta), 
\ee
and $v_z = -\frac{\kappa}{k}aJ_0(\kappa r)\sin(kz-\omega t+\theta).$
In either case the final result is 
\be \label{helicity3}
\textrm{curl}\mathbf{v} = \mp \sqrt{k^2 + \kappa^2}\mathbf{v}, \quad \textrm{and} \quad
\mathbf{v} \cdot \textrm{curl}\mathbf{v} = \mp \sqrt{k^2 + \kappa^2}|\mathbf{v}|^2. 
\ee
When the liquid is contained in a solid cylinder of radius $b$, $\omega$ is replaced by Eq. \rr{omega}
and $\kappa$ by $\gamma_n/b$, where $\gamma_n$ is the $n$-th zero of $J_1(x)$.

%
%enstrophy-type integrals do exist for all
%even-dimensional ideal fluid flows, and so do helicity-type integrals for all odddimensional
%flows Arnold p. 43

\section{\label{sec: modified}Modified Taylor-Proudman Theorem}
For simplicity we consider Cartesian coordinates. The ``geostrophic'' form (i.e. Navier-Sokes with odd viscosity, without inertia and without shear viscosity) of the equations is 
\be \label{TP1}
\frac{1}{\rho}\frac{\partial p}{\partial x} = -\nu_o \nabla^2_2 v, \quad \frac{1}{\rho}\frac{\partial p}{\partial y} = \nu_o \nabla^2_2 u, \quad \frac{\partial p}{\partial z} = 0,  
\ee
where $\nabla^2_2 = \partial_x^2 + \partial_y^2$. This reduction is possible by invoking the 
requirement $u \ll \nu_o /\ell$ where $u$ and $\ell$ are characteristic velocity
and length scales respectively. This inequality can be derived by balancing the inertial terms
$\mathbf{v}\cdot \nabla \mathbf{v} \sim u^2/\ell$ with the odd viscous term $\nu_o \nabla_2^2 u\sim \nu_o u/\ell^2$ and requiring that latter is dominant.

Differentiating the first two with respect to $z$ and considering the third we obtain
\be
\nabla^2_2 \partial_z u =0, \quad \textrm{and} \quad \nabla^2_2 \partial_z v =0. 
\ee
Eliminating the pressure by cross differentiation of the fist two equations we obtain
\be
\nabla^2_2 (\partial_x u + \partial_y v) =0.
\ee
Thus, from continuity we also obtain 
\be
\nabla_2^2 \partial_z w =0.
\ee
Defining
\be \label{UVW}
(U,V, W) = \ell^2\nabla^2_2(u,v,w),
\ee
(the length-scale is determined, for instance from the size of the vessel) we derive the Taylor-Proudman theorem for the velocity field $U,V,W$, that is 
\be \label{mTP}
\partial_z U = \partial_z V = \partial_zW =0
\ee
and this can be considered as a liquid in a frame rotating with angular velocity $\Omega = \nu_o\ell^{-2}$. 
Thus, when the odd viscosity terms are larger than inertia, there is a superposition of a two-dimensional
motion in the lateral ($x-y$) plane and a vertical motion, independent of $z$.

Some familiar behavior at a boundary can also be recovered. Because of the no-penetration
condition $w=0$ on a solid boundary we have $W = 0$ on the same boundary ($(\partial_x^2 + \partial_y^2)w$ must be zero
on the boundary). Thus, when 
a streamline parallel to the axis meets a stationary boundary, this implies that $W$ is zero everywhere.

For axisymmetric systems a further simplification takes place. letting $V_r= \mathcal{L} v_r $ and $V_\phi = \mathcal{L} v_\phi$ where $\mathcal{L}$ was defined in Eq. \rr{L}, equations \rr{TP1} become 
\be
\nu_o V_\phi = -\frac{1}{\rho}\frac{\partial p}{\partial r} \quad \textrm{and} \quad - \nu_o V_r =0.
\ee
Thus, $V_r$ is zero everywhere and the flow proceeds in spirals. With $V_z= \mathcal{L} v_z $
the continuity equation becomes $\partial_r V_r + \partial_z V_z =0$. Thus, 
$\partial_zV_z=0$ everywhere, giving rise to the Taylor-Proudman theorem.  

%Since $\frac{1}{r}\partial_r(rv_r) + \partial_z v_z=0$, letting $V_r= \mathcal{L} v_r $ and $V_z = \mathcal{L} v_z$
%implies that $V_r =0$ everywhere. 
%Thus, $v_r$ satisfies $r^2\partial_r^2 v_r + r \partial_r v_r - v_r =0$, so $v_r = Ar + B/r$
%(this can also be seen by the governing equation \rr{nsphi}). 

\begin{figure}
%   \centering
    \begin{subfigure}[t]{0.5\textwidth}
%        \centering
        \includegraphics[width=\linewidth]{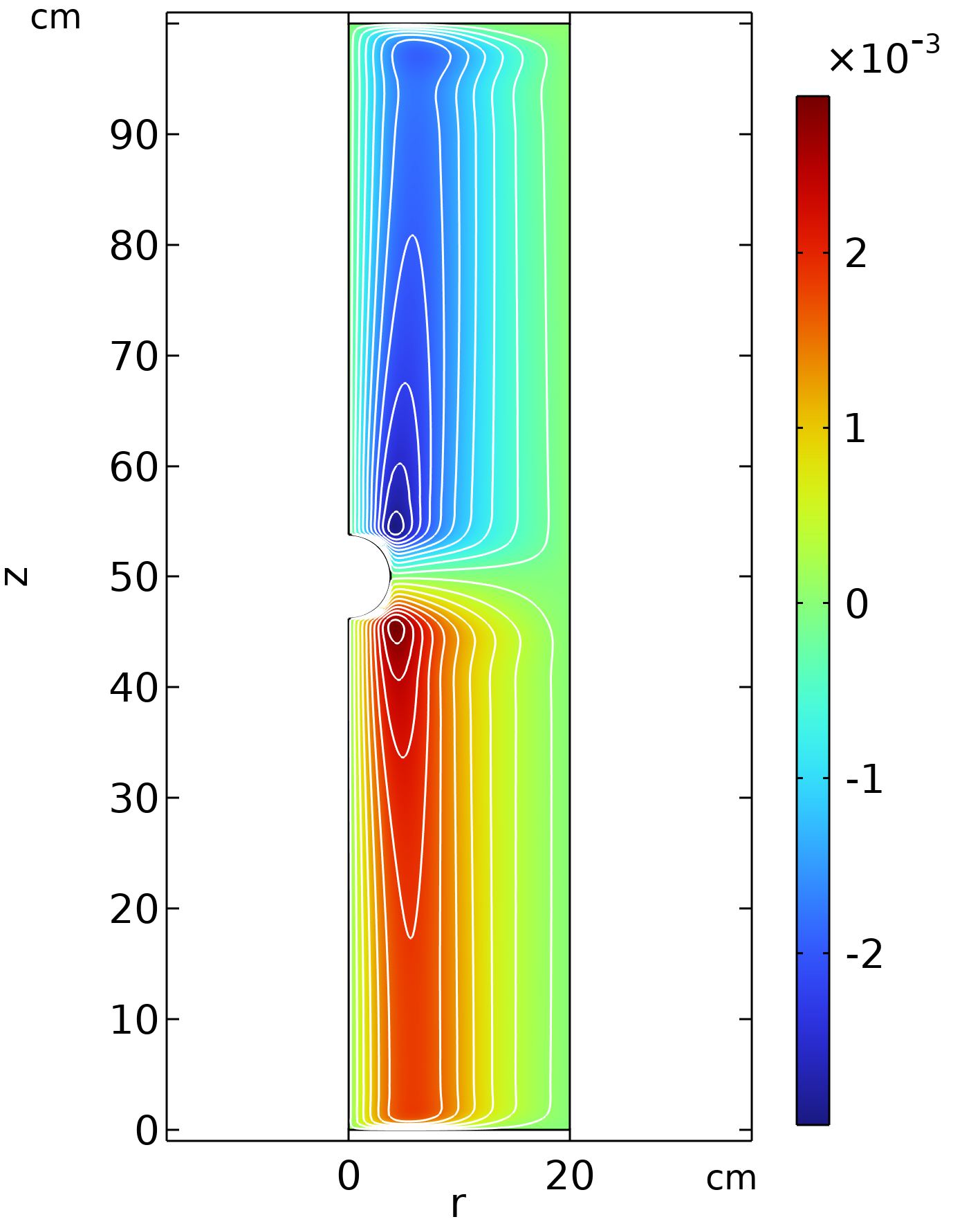} 
%        \caption{} \label{chiraltorque1}
    \end{subfigure}
%    \hfill
    \begin{subfigure}[t]{0.5\textwidth}
%       \centering
        \includegraphics[width=\linewidth]{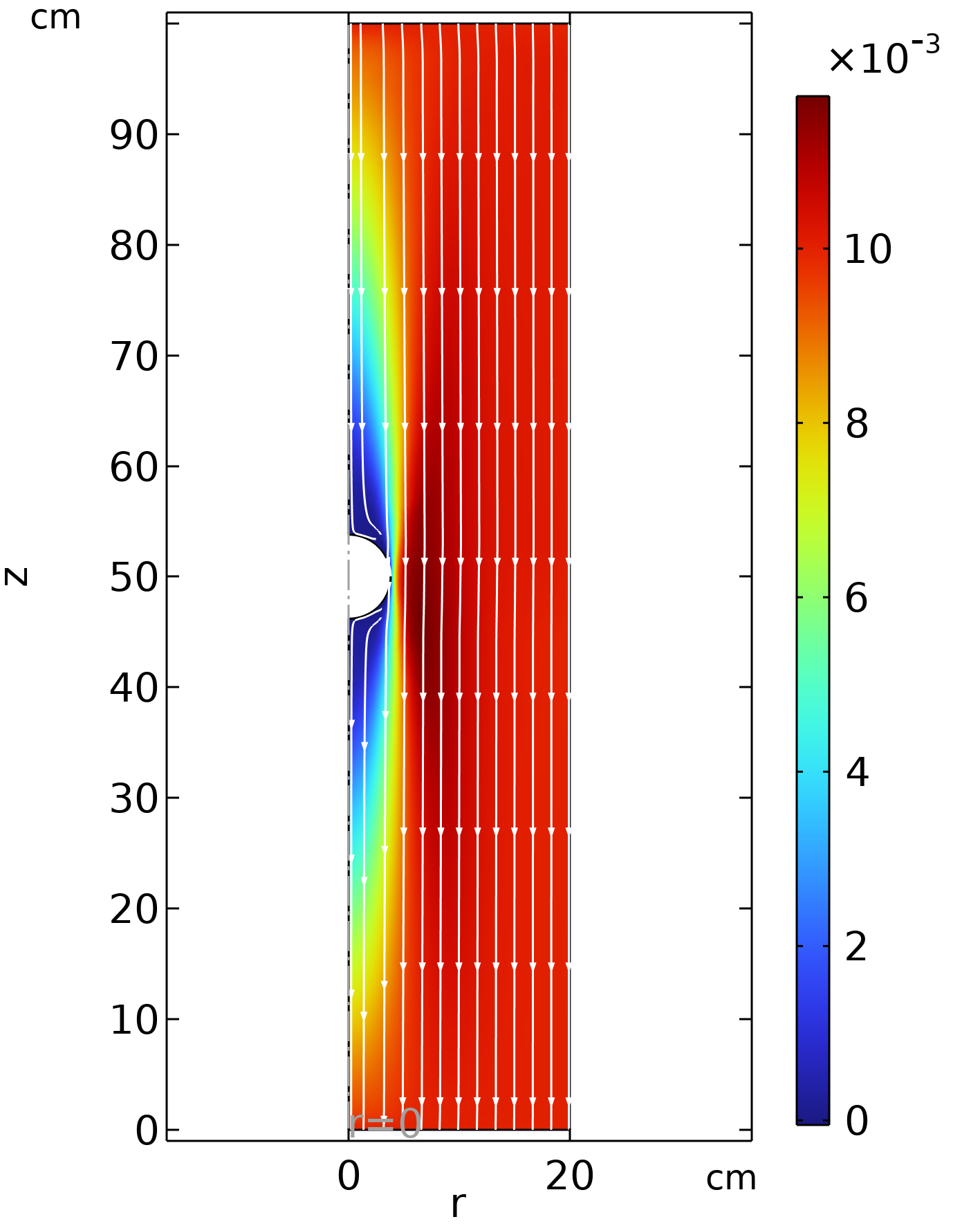} 
%        \caption{Spatial distribution of the observables $v_p, \sigma^{\textrm{ch}}$ and $v^{\textrm{ch}}$.} \label{vpschvchN}
    \end{subfigure}
     %  \vspace{1cm}
%    \begin{subfigure}[t]{\textwidth}
%    \centering
%        \includegraphics[width=\linewidth]{example-image-c.pdf} 
%        \caption{Price regulation} \label{fig:timing3}
%    \end{subfigure}
    \caption{\label{velocity2Dazimaxial}Distribution of azimuthal (left panel) and axial velocity (right panel) in an odd
    liquid moving slowly and meeting an immobile sphere (of radius $3.8$ cm) located at elevation $z=50$ cm 
 at the center axis of a cylinder. 
   Liquid enters from the top ($z=100$ cm) and exits at the bottom ($z=0$). The sphere is not allowed to rotate. Counter-rotation of liquid takes place above and below
    the sphere in the azimuthal direction (this was also observed by \citet[Fig.4(c)]{Khain2022} for the analogous problem in Stokes flow).
    A Taylor-type column is also visible in the right panel, symmetrically placed above and
    below the sphere} 
\end{figure}

\section{\label{sec: Taylor}Taylor columns in an odd viscous liquid}
The Navier-Stokes equations \rr{TP1} written in the form
\be \label{TP2}
\frac{1}{\rho}\frac{\partial p}{\partial x} = -\frac{\nu_o}{L^2} V, \quad \frac{1}{\rho}\frac{\partial p}{\partial y} = \frac{\nu_o}{L^2} U, \quad \frac{\partial p}{\partial z} = 0,  
\ee
by employing the definition of $U,V$ and $W$ in Eq. \rr{UVW}, are suggestive of the existence of Taylor columns in an odd viscous liquid. By this we mean that when 
an axisymmetric body moves slowly in an odd viscous liquid, a column or a ``slug'' will be pushed ahead
of the body with zero axial velocity relative to the body. 
In the inviscid limit implied by Eq.\rr{TP2} (meaning the shear viscosity is zero), the column is a cylinder but it will be modified
by the presence of shear viscosity. A rear slug will also be present. In general the motion of the liquid
in the slug can not be determined from the simple equations \rr{TP2}. In reality a number of boundary
layers exist which act as a conduit that transports liquid between different locations. 

The determination of the flow structure is not only difficult but also changes dramatically when one
alters the parameters and the geometry. Taylor columns were studied in the past in the context 
of slowly moving bodies immersed in a liquid rotating rigidly, see for instance the following
comprehensive references
\citep{Moore1968, Maxworthy1970, Tanzosh1994, Bush1994} and references therein.

We can define some useful dimensionless parameters for the odd viscosity dominated problems, such as the Taylor $\mathcal{T}$ and Maxworthy $\mathcal{M}$ numbers (\citet{Maxworthy1970} employed the notation
$N$ to denote the rotating counterpart of the latter)
\be \label{MaxworthyM}
\mathcal{T} = \frac{\nu_o}{\nu_e}, \quad \mathcal{M} = \frac{\nu_o}{aU}. 
\ee
The Taylor number can be understood as an inverse Ekman number denoting the strength of odd to even viscosity (angular velocity to even viscosity in the rotating fluid case) and the Maxworthy number is the 
ratio of odd viscous force  to inertial forces or the odd viscosity-induced inertial wave propagation velocity 
to convection velocity.

\begin{figure}
\vspace{-5pt}
\begin{center}
\includegraphics[height=3.4in,width=3.6in]{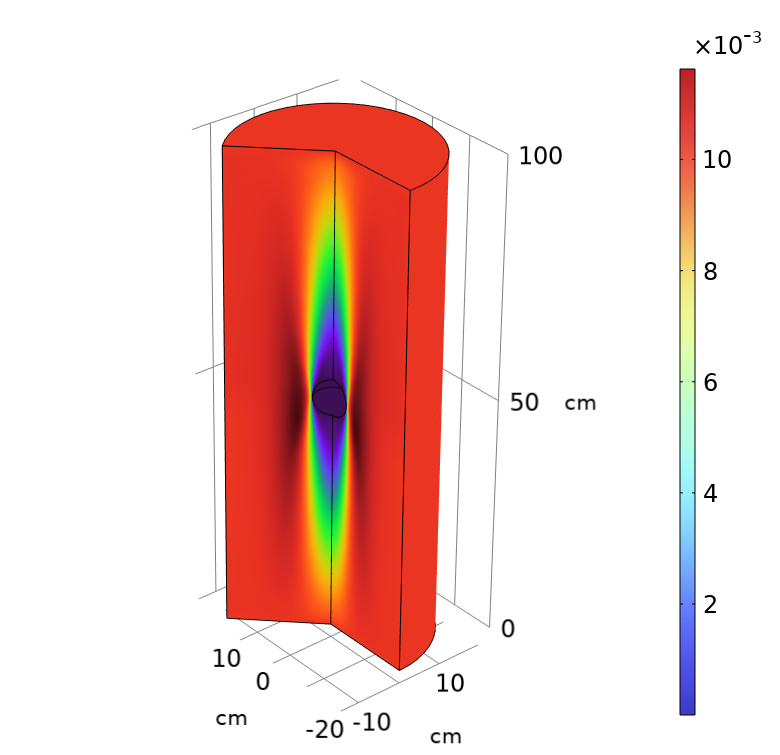}
\vspace{-0pt}
\end{center}
\caption{
\label{velocity3D}
{Three-dimensional realization of odd viscous flow around an immobile sphere (of radius $3.8$ cm) located at elevation $z=50$ in a moving cylinder with liquid entering from the top ($z=100$ cm) and exiting at the bottom ($z=0$), with the cylinder speed (the sphere is not allowed to rotate). Color-bar denotes the strength of the axial velocity $w$.  
A Taylor column of low axial velocity is visible circumscribing 
the sphere and surrounding the central axis. Parameters employed in to produce this figure:
Odd viscosity coefficent $\eta_o = 0.5$ g/(cm sec), shear viscosity $\eta = 0.01$ g/(cm sec), 
cylinder radius $20$ cm, sphere radius $3.8$ cm, cylinder height $H=100$ cm, liquid density
$\rho = 1 \textrm{g/cm}^3$, liquid velocity in the $-\hat{\mathbf{z}}$ direction $U=0.01$ cm/sec. 
These give a Taylor number based on the sphere length-scale $\mathcal{T} = 50$ and 
Maxworthy number $\mathcal{M} = 26.316$, see Eq. \rr{MaxworthyM}. }}
\vspace{-0pt}
\end{figure}

{
We can compare the Maxworthy number $\mathcal{M}$ in \rr{MaxworthyM} for an odd viscous liquid to the inverse Rossby number $\textrm{Ro}^{-1} = \frac{\Omega a}{U}$ in 
\rr{RoEk} when the parameters $\nu_o$ and $\Omega$ are held constant. It is clear that Taylor columns
in an odd viscous liquid are favored at small ({in-plane}) length scales, while Taylor columns 
are favored at large length-scales in 
rigidly-rotating liquids. This is clear due to the fact that odd viscosity multiplies second order spatial
derivatives. Even if its value is small and unimportant in general, 
observable effects will be present close to boundaries, such 
as sharp boundary-layers. 
Thus, one should consider odd viscous effects (described by the specific constitutive law \rr{sigma0}) under the restrictions posed by this discussion, which might limit the applicability
of the odd viscous liquids in comparison to their (non-odd viscous) rotating counterparts. 
}

The pressure $p$ is a streamfunction and thus constant on a streamline of the flow $(U,V,W)$.
A finite-length cylinder with generators parallel to the rotating axis and moving horizontally in a rotating liquid will have a liquid velocity parallel to its 
generators and a column will accompany its motion \cite[\S 12.2]{Yih1988}. 
Likewise, a solid body translating slowly along the axis of the cylinder will be accompanied by a
column of fluid with generators parallel to the axis.

Figures \ref{velocity2Dazimaxial} and \ref{velocity3D} display the salient features of Taylor columns
in odd viscous liquids. Liquid flow entering from the top of a cylinder encounters an immobile 
sphere located at the central axis. The two ``slugs'' located above and below the sphere
in Figures \ref{velocity2Dazimaxial} (right panel) and \ref{velocity3D} are characterized by
the sharp blue color. There is a swirling flow that takes place above and below the sphere
with opposite sense of rotation, Figure \ref{velocity2Dazimaxial} (left panel). 
This is further discussed below.

\begin{figure}
   % \centering
    \begin{subfigure}[t]{0.5\textwidth}
       % \centering
        \includegraphics[width=\linewidth]{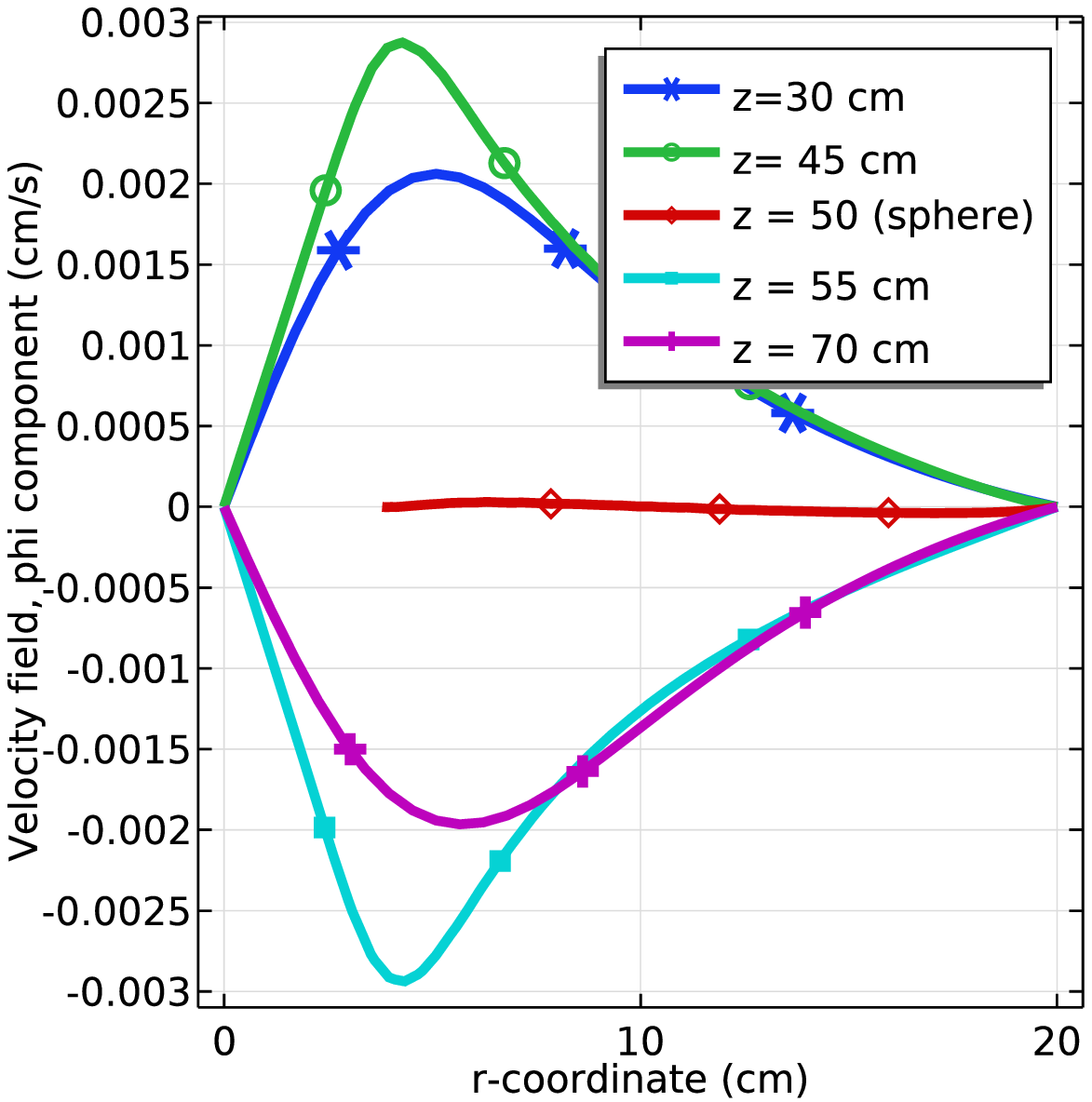} 
%        \caption{} \label{chiraltorque1}
    \end{subfigure}
  %  \hfill
    \begin{subfigure}[t]{0.5\textwidth}
       % \centering
        \includegraphics[width=\linewidth]{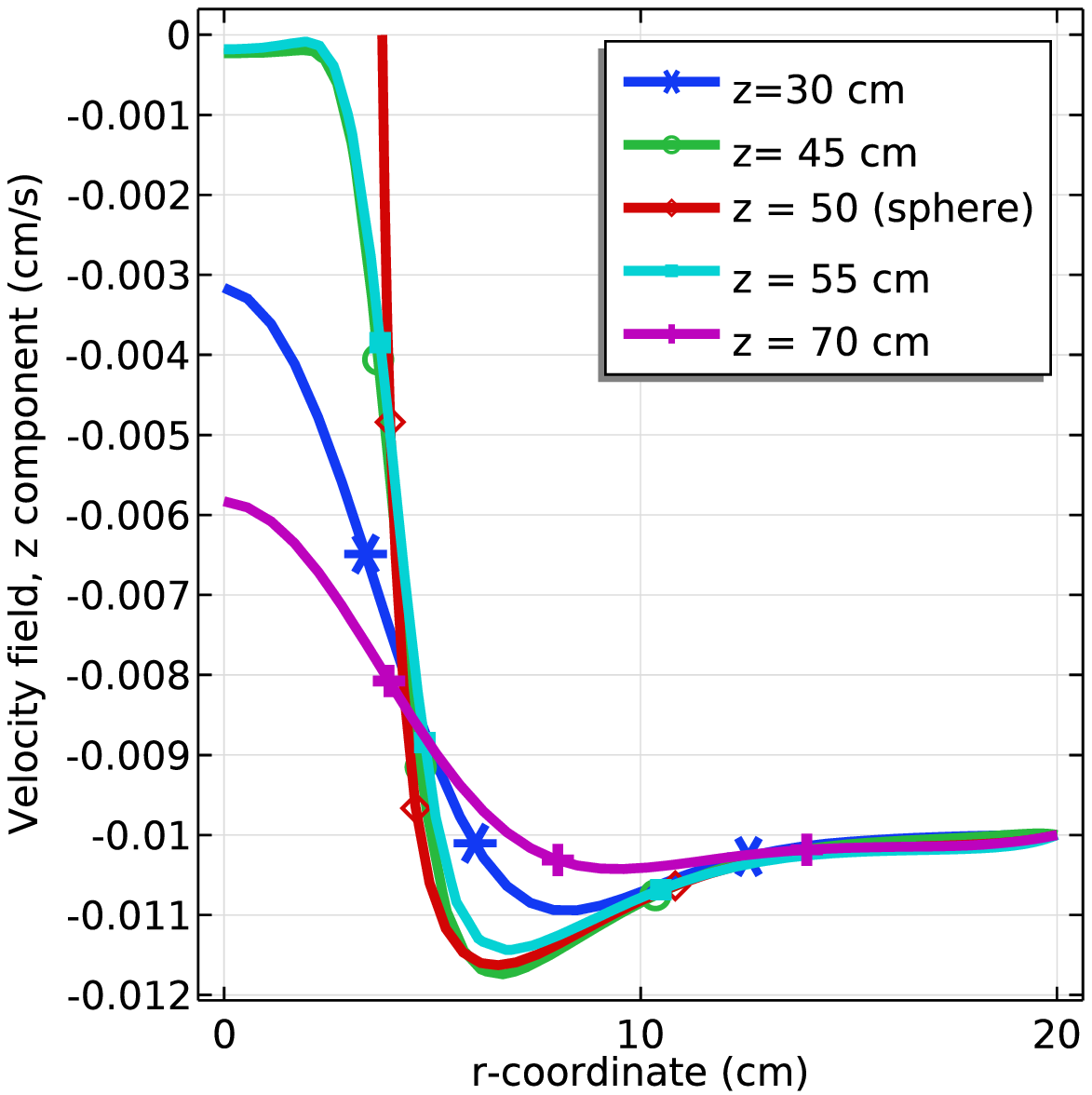} 
%        \caption{Spatial distribution of the observables $v_p, \sigma^{\textrm{ch}}$ and $v^{\textrm{ch}}$.} \label{vpschvchN}
    \end{subfigure}
     %  \vspace{1cm}
%    \begin{subfigure}[t]{\textwidth}
%    \centering
%        \includegraphics[width=\linewidth]{example-image-c.pdf} 
%        \caption{Price regulation} \label{fig:timing3}
%    \end{subfigure}
    \caption{\label{velocityazimaxial}{  Plots of azimuthal velocity $v_\phi$ (left panel) and axial velocity $v_r$ (right panel) vs. distance $r$ from the central axis of the cylinder of Fig. \ref{velocity3D}, 
    for an odd viscous liquid
    past a stationary sphere of radius $3.8$ cm and located at elevation $z=50$ cm.
    Each curve corresponds to a specific elevation in the cylinder where a probe has been placed. Two probes are located below the sphere at $z=30$ and 
    $z=45$ cm, one lies at the level of the sphere ($z=50$ cm) and two above the sphere 
    at $z=55$ and $z=70$ cm.     } 
The left panel shows the counter-rotating azimuthal velocity components above and below the sphere as was also depicted in the left panel of Fig. \ref{velocity2Dazimaxial}. The magnitude of the azimuthal velocity decreases as one approaches the cylinder lids (at $z=0$ and $z= 100$ cm) because liquid entering or leaving
the cylinder has been set to have a vanishing azimuthal velocity component. The axial
    velocity of the liquid in the right panel shows similarities (and differences) with respect to the analogous
    flow of a fluid in a rotating cylinder as this was clearly described by \citet[Fig.7]{Maxworthy1970}}
\end{figure}

The exact form of a flow associated with Taylor columns induced by a solid body 
is a complicated problem depending on parameter regimes (Ekman, Reynolds and Rossby numbers) as 
well as the geometry (finite or infinite cylinder), geometry of the body and ultimately, its constitution (whether
it is a solid or a liquid). 
Here we will not pause to carry-out a detailed enumeration of special cases arising in the various
parameter regimes, geometries and materials; we will only point-out certain qualitative similarities that
exist between an odd viscous liquid and rigidly rotating flow. 

Figure \ref{velocityazimaxial} displays azimuthal and axial velocities of the liquid flowing in the cylinder 
of Figures \ref{velocity2Dazimaxial} and \ref{velocity3D}. Probes located at different elevations of 
the cylinder measure velocities as they vary in radial direction, from the cylinder central axis to its
external surface, with a view to compare our results to the experiments of \citet{Maxworthy1970}.
In these numerical simulations the Taylor and Maxworthy numbers are $\mathcal{T} = 50$ and 
$\mathcal{M} = 26$, respectively.

We find a velocity defect region (at $z=70$ cm, cf. 
region I in the experiments of \citep{Maxworthy1970}) that is, the region where the velocity 
is less than the free stream one. This is surrounded by the region $5<r<15$, which has a velocity
in excess of the free stream (region V in the experiments of \citep{Maxworthy1970}). 
The flow at $z=30$ cm, is narrower and has larger axial velocity than its mirror counterpart (at $z=70$ cm).
Its angular velocity seems to be of the same magnitude compared with its mirror image (and this 
differs from the flow character in region VII of \citep{Maxworthy1970}). 

The probe at $z=55$ cm has a zero axial velocity close to the axis as this is determined by the 
Taylor-Proudman theorem, that the axial velocity of the slug is the same as the velocity of the body. 
The swirl however is nonzero and is actually quite large. This is region II in the experiments of 
\citep{Maxworthy1970}. An Ekman layer induces a slow axial velocity where the surface
of the sphere meets the anisotropy axis (close to $z=45$ cm probe, and near $r=0$). A
sharp Ekman boundary-layer is seen to form at $z=50$ cm, $r\sim 4$ cm.

We have not observed an oscillatory region downstream the sphere (region III of \citep{Maxworthy1970})
and we have not determined whether a region of clear fluid exists downstream and adjacent to the 
anisotropy axis.

The left panels of Figures \ref{velocity2Dazimaxial} and \ref{velocityazimaxial}
show the counterotating character of the flow above and below the sphere, similar to 
Fig. 1 of \cite{Moore1968}. 
The above discussion displays an indirect verification for the validity of the Taylor-Proudman theorem in the case of odd viscous flow with a velocity field $(v_r,v_\phi,v_z)$. $v_z$ has the 
velocity of the body ($v_z=0$), $v_r$ is virtually zero and the flow outside the column is independent of $z$.

\section{\label{sec: vortex}Vortex stretching and vortex twisting in a three-dimensional odd-viscous liquid. }
The constitutive law \rr{sigma0} can be written in Cartesian coordinates in the form
\be \label{sigma1}
\bm{\sigma}' = \eta_o 
\left(\begin{array}{ccc}
-\left(\partial_x v + \partial_y u \right) &  \partial_x u  - \partial_y v & 0\\
 \partial_x u  - \partial_y v & \partial_x v + \partial_y u   & 0\\
0&0&0
\end{array}
\right).
\ee
It is possible to define a modified pressure $\tilde{p} = p + \eta_o \zeta$ where 
$\zeta = \partial_x v - \partial_y u$ here
is the component of vorticity in the $z$ direction.   
Then, the odd Navier-Stokes equations are 
\be \label{Duvw}
\rho \frac{D u}{Dt}  = - \partial_x \tilde{p} + \eta_o \partial_y (\partial_z w), \quad 
\rho \frac{D v}{Dt}  = - \partial_y \tilde{p} - \eta_o \partial_x (\partial_z w), \quad 
\rho \frac{D w}{Dt}  = - \partial_z \tilde{p} + \eta_o \partial_z \zeta. 
\ee
What this equation shows is that vortex stretching $\partial_z w$ will be important on a region in the $x-y$ plane
with vorticity $\zeta$. To show this, let $\textrm{curl} \mathbf{v} = (\xi, \eta, \zeta)$ be the components of vorticity in Cartesian coordinates and 
consider a fluid particle whose vorticity points in the $z$ direction instantaneously (we perform this 
to simplify the nonlinear term $\textrm{curl} \mathbf{u} \cdot \nabla \mathbf{v}$ that arises in the vorticity equation \rr{dtcurlvnl}). 
Taking the curl of \rr{Duvw}, or considering directly the vorticity equation \rr{dtcurlvnl} we obtain
\be \label{xietazeta}
\frac{D \xi}{Dt} = (\zeta - \eta_o\nabla_2^2) \partial_z u, \quad\frac{D \eta}{Dt} = (\zeta - \eta_o\nabla_2^2) \partial_z v, \quad\frac{D \zeta}{Dt} = (\zeta - \eta_o\nabla_2^2) \partial_z w. 
\ee
Thus, the well-known vortex twisting, represented by the quantities $\partial_zu$ and $\partial_zv$
(cf. \citep[\S 6.6]{Tritton1988})
is now enhanced by the extra term $\eta_o (k_x^2 + k_y^2)$ appearing in the round brackets 
in Eq. \rr{xietazeta} induced by odd viscosity. 
Likewise, vortex stretching, represented by the quantity $\partial_z w$ is also enhanced
by odd viscosity. 

Of course, for a two-dimensional incompressible odd viscous liquid where the velocity does not depend on $z$, we have
\be
\frac{D \zeta}{Dt} = 0,
\ee
as is also known from the ``absorption'' of the odd viscous force density into the pressure gradient, see eg. 
\citep{Ganeshan2017}, and thus the vorticity of a fluid particle in a two-dimensional odd viscous liquid is conserved.

\section{\label{sec: eta4}The effect of the $\eta_4$ odd viscous stress tensor}
{Referring to the stress tensor \rr{seta1234}, in this article we tacitly assumed that only the coefficient
$\eta_3$ was nonvanishing which led the odd stress tensor to acquire the form \rr{sigma1} 
in Cartesian coordinates. The coefficient $\eta_4$ in \rr{seta1234} gives rise however to an additional
odd stress tensor. In this section we discuss the consequences of the latter, in the context of the effects developed in this article. 

As in section \ref{sec: hydrodynamics} we consider the field $\mathbf{b}$ to lie in the $z$-direction. 
Then, the odd stress tensor acquires the additional components 
\be \label{sigma4}
\bm{\sigma}' = \eta_4 
\left(\begin{array}{ccc}
0 & 0  & -(\partial_y w + \partial_z v)\\
0 & 0  & \partial_x w + \partial_z u\\
 -(\partial_y w + \partial_z v)&\partial_x w + \partial_z u&0
\end{array}
\right), 
\ee
where we adopted the opposite sign to \citep[Eq. (58.16)]{Landau1981}.
We proceed below to examine the form of inertial-like waves in an odd viscous liquid whose constitutive
law is the combination of \rr{sigma1} and \rr{sigma4}.

\subsection{Plane-polarized waves}
We define the linear operator
\be \label{Leta4}
\mathcal{S} = (\nu_o-\nu_4)\nabla^2_2 + \nu_4\partial_z^2, 
\ee
where $\nu_4 = \eta_4/\rho$ and $\nabla^2_2 = \partial_x^2 + \partial_y^2$. 
To obtain an understanding of the effect odd viscosity parameters have on the type of solutions we 
classify the operator $\mathcal{S}$ in \rr{Leta4} as follows
\begin{itemize}
\item $\mathcal{S}$ is elliptic when $\nu_o>\nu_4$.
\item $\mathcal{S}$ is hyperbolic when $\nu_o<\nu_4$.
\item $\mathcal{S}$ is parabolic when $\nu_o=\nu_4$.
\end{itemize}
Here we have assumed that $z$ plays the role of the time-like variable. This is the standard route
followed in rigidly-rotating liquids, see \citep[\S 12.6]{Whitham1974}.

With the notation $\zeta = \partial_x v - \partial_y u$ for the component of vorticity in the $z$ direction and a modified pressure $\tilde{p} = p + \eta_4 \zeta$,
the Navier-Stokes equations \rr{DvDt} are replaced by 
\be \label{DvDt2}
\frac{D\mathbf{v}}{Dt} = -\frac{1}{\rho}\nabla \tilde{p} +\mathcal{S}\hat{ \mathbf{z}} \times \mathbf{v},
\ee
and the vorticity equation \rr{dtcurlv} by
\be \label{dtcurlv2}
\partial_t\textrm{curl} \mathbf{v} = - \mathcal{S} \frac{\partial \mathbf{v}}{\partial z}. 
\ee
\begin{figure} 
\begin{center}
\includegraphics[height=3in,width=3in]{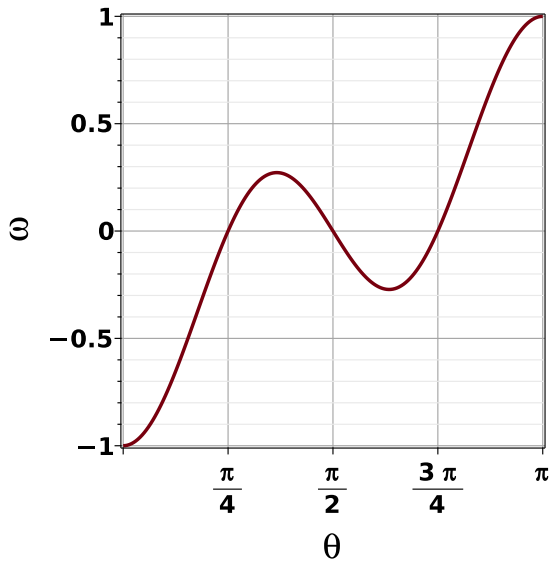}
\end{center}
\vspace{-5pt}
\caption{{Plane-polarized wave dispersion $\omega$ (Eq. \rr{dispeta4}) versus angle $\theta$ between the wavevector $\mathbf{k}$
and the $z$- (anisotropy) axis (setting $k=-\nu_4 = 1, \nu_o = 0$). It differs qualitatively from its
$\eta_o \neq 0$ counterpart displayed in Fig. \ref{omega1} by crossing the $\omega=0$ axis at 
angles different that $\pi/2$ and by displaying additional inflection points which signal that the maxima
of group velocity may appear at angles that differ to $\pi/2$. 
}
\label{omegaeta4} }
\end{figure}
With $
\mathbf{v} = \mathbf{A} e^{i(\mathbf{k}\cdot \mathbf{r} - \omega t)} 
$ the vorticity equation \rr{dtcurlv2} becomes a system of three equations for the three unknown
components of the amplitude $\mathbf{A}$. This system has a nontrivial solution when the 
determinant of the matrix 
\be \label{system4}
\left(\begin{array}{ccc}
-ik_z\mathcal{S}(\mathbf{k}) & \omega k_z  & -\omega k_y\\
-\omega k_z &-ik_z\mathcal{S}(\mathbf{k}) & \omega k_x\\
\omega k_y&-\omega k_x&-ik_z\mathcal{S}(\mathbf{k})
\end{array}
\right)
\ee
vanishes. Here, 
\be \label{Leta4a}
\mathcal{S}(\mathbf{k}) = - (\nu_o - \nu_4) (k_x^2+k_y^2) - \nu_4 k_z^2. 
\ee

The dispersion relation becomes
\be \label{dispeta4}
\omega = \mp \mathcal{S}(\mathbf{k})\frac{k_z}{k} \quad \textrm{or} \quad 
\omega = \pm \cos \! \left(\theta \right) k^{2} \left[\nu_o -\nu_4 -\left(\nu_o -2 \nu_4 \right)\cos^{2}\theta  \right]. 
\ee
Here and below the reader can keep in mind the parabolic case $\nu_o=\nu_4$ and the elliptic case
$\nu_o = 2\nu_4$ (parabolic and elliptic with respect to the operator $\mathcal{S}$ in \rr{Leta4}) which
simplify all relations significantly and are to be discussed in what follows. 

We display the dispersion Eq. \rr{dispeta4} for $\nu_o = 0$
in Fig. \ref{omegaeta4}, to be compared with Fig. \ref{omega1}. 
This dispersion relation is interesting as it crosses the $\omega=0$ axis at angles $\theta = \pi/4$ (different
to the $\pi/2$ of rigidly-rotating liquids or the $\eta_o$ odd viscous liquid). Likewise, it has an inflection
point at $\theta \neq \pi/2$, which signals the presence of a maximum for the group velocity that 
differs to the ones of rigidly-rotating liquids and the $\eta_o$ odd viscous liquid. 
The group velocity \rr{polgroup1} is replaced by
\begin{align} \label{cgthxyzeta4}
\left(\frac{\partial \omega}{\partial k_x}, \frac{\partial \omega}{\partial k_y} \right)
&= \pm k \left(\left(\nu_o -2 \nu_4 \right) \cos^{2}\theta+\nu_o -\nu_4 \right)\sin \theta  \cos \theta( \cos \phi, \sin \phi),
\\
\label{cgthxyzeta4b}
\frac{\partial \omega}{\partial k_z} &= \pm k \left(\left(\nu_o -2 \nu_4 \right) \cos^{4}\theta +\left(-2 \nu_o +5 \nu_4 \right)\cos^{2}\theta+\nu_o -\nu_4 \right), 
\end{align}
with modulus 
\begin{align}
|c_g| = &k \left[-5 \left(\nu_o -2 \nu_4 \right)^{2} \cos^{6}\theta + \left(7\nu_o -16 \nu_4\right) \left(\nu_o -2 \nu_4 \right) \cos^{4}\theta \right. \\
& \left.
-3 \left(\nu_o -\nu_4 \right) \left(\nu_o -3 \nu_4 \right) \cos^{2}\theta+\left(\nu_o -\nu_4 \right)^{2}\right]^{1/2}.
\end{align}
Its components and modulus are displayed in Fig. \ref{cgeta4} for the case $\nu_o =0$. The modulus of the group
velocity displays a maximum at $\theta \neq \pi/2$ as expected from the properties of the corresponding
dispersion relation \rr{dispeta4}. 
\begin{figure} 
\begin{center}
\includegraphics[height=3in,width=3in]{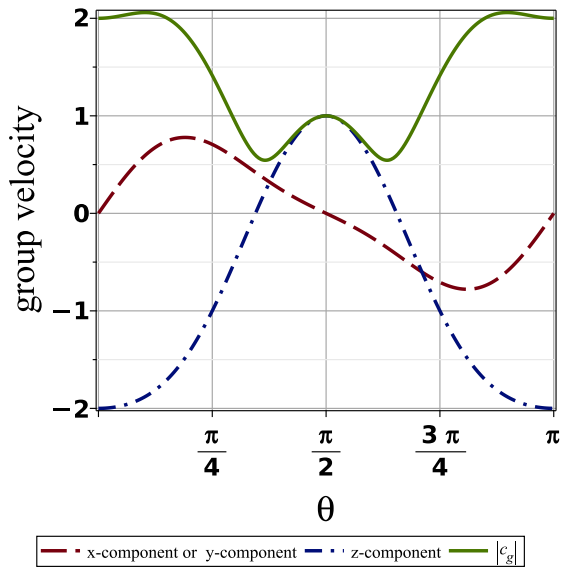}
\end{center}
\vspace{-5pt}
\caption{{Group velocity components \rr{cgthxyzeta4} and magnitude versus angle $\theta$ between the wavevector $\mathbf{k}$
and the $z$- (anisotropy) axis, setting $k=-\nu_4 = 1, \nu_o = 0$ (and $ \phi =\pi/4$, for simplicity). 
The magnitude develops maxima at angles $\theta \neq \pi/2$, differing to those of the rigidly-rotating liquid
or the $\eta_o$ odd viscous liquid, cf. Fig. \ref{cg1}.} 
\label{cgeta4} }
\end{figure}

When the flow field is determined by the plane-polarized waves  $
\mathbf{v} = \mathbf{A} e^{i(\mathbf{k}\cdot \mathbf{r} - \omega t)} 
$
considered in this section, its helicity is conserved for an odd viscous liquid that incorporates both 
constitutive laws \rr{sigma1} and \rr{sigma4}. That is the case because in wave-number and 
frequency-space the linearized vorticity
equation \rr{dtcurlv2} can be written in the form
\be
-i\omega  \mathbf{B} =- i \mathbf{A}k_z \mathcal{S}(\mathbf{k}),
\ee
when $\textrm{curl} \mathbf{v} = \mathbf{B} e^{i(\mathbf{k}\cdot \mathbf{r} - \omega t)}$ 
and $\mathcal{S}(\mathbf{k})$ is defined in Eq. \rr{Leta4a}. 
Since $\omega = \mp \mathcal{S}(\mathbf{k})\frac{k_z}{k}$ (from \rr{dispeta4}) we obtain $\mathbf{B} = \mp k \mathbf{A}$, or 
\be
\textrm{curl} \mathbf{v} = \mp k\mathbf{v}. 
\ee
Thus, the helicity of the flow field determined by the odd stress tensors \rr{sigma1} and \rr{sigma4} is conserved
\be \label{helicityeta4}
\mathbf{v} \cdot \textrm{curl}\mathbf{v} = \mp k|\mathbf{v}|^2. 
\ee
In addition, $\mathbf{c}_g \cdot \mathbf{c}_p  = 2
\left[ (\nu_o-2\nu_4)\cos^2\theta -\nu_o+\nu_4 \right]^2k^2\cos^2\theta= 2|\mathbf{c}_p|^2$, 
so the last two lines of Table \ref{table:table1} remain unchanged.

Equation \rr{DvDt2} can be employed to investigate the possibility of Taylor column formation in an 
odd viscous liquid that incorporates the stress tensor \rr{sigma1} and \rr{sigma4}. Dropping the left-hand side of \rr{DvDt2}, 
it reads, in component form
\be \label{TPeta4}
\frac{1}{\rho}\frac{\partial \tilde{p}}{\partial x} = -\mathcal{S} v, \quad \frac{1}{\rho}\frac{\partial \tilde{p}}{\partial y} = \mathcal{S} u, \quad \frac{\partial \tilde{p}}{\partial z} = 0,  
\ee
and carrying-out the same manipulations as in section \ref{sec: modified} we obtain the conditions
\be \label{condTPeta4}
\partial_z\mathcal{S} u = \partial_z\mathcal{S}v=\partial_z\mathcal{S}w =0, \quad \textrm{and} \quad \partial_x\mathcal{S}u +\partial_y\mathcal{S} v =0,
\ee
which replace \rr{mTP}. The form of both equations \rr{TPeta4} and \rr{condTPeta4} is appealing
and resembles \rr{TP1} and \rr{mTP}, respectively. They still lead to Taylor column-like structures 
which however are not identical to those of rigidly-rotating liquids.

In Fig. \ref{velocity2Dazimaxialeta4} we repeat the simulations of Fig. \ref{velocity2Dazimaxial}
now with vanishing odd viscosity $\eta_o$ and $\eta_4 = 1$ g/(cm sec). 
On the right panel of Fig. \ref{velocity2Dazimaxialeta4} we observe a column circumscribing the sphere. 
It does not have the characteristics of a Taylor column present in rigidly-rotating liquids (where $w$ is
equal to the velocity of the body circumscribed by 
the column). Here the variation of $w$ with respect to $z$, away from boundaries, is mostly
of the form of a lower degree polynomial in $z$ plus some oscillatory behavior of low amplitude.   
This behavior is expected from the form of governing equation $\partial_z\mathcal{S}w =0$ in 
\rr{condTPeta4}. Its particular solution
is some quadratic polynomial (at most) in $z$ to which one can superpose the solution of the homogeneous problem which is oscillatory
in $z$.

The left 
panel of Fig. \ref{velocity2Dazimaxialeta4} shows additional lobes of azimuthal velocity. This 
agrees with the observations of \citep[Fig.4(d)]{Khain2022} for the analogous problem in Stokes flow. 
There are regions where the sense of rotation changes with distance from 
the axis, i.e. the azimuthal velocity lobes that are additional to the ones of Fig. \ref{velocity2Dazimaxial}.

\begin{figure}
%   \centering
    \begin{subfigure}[t]{0.5\textwidth}
%        \centering
        \includegraphics[width=\linewidth]{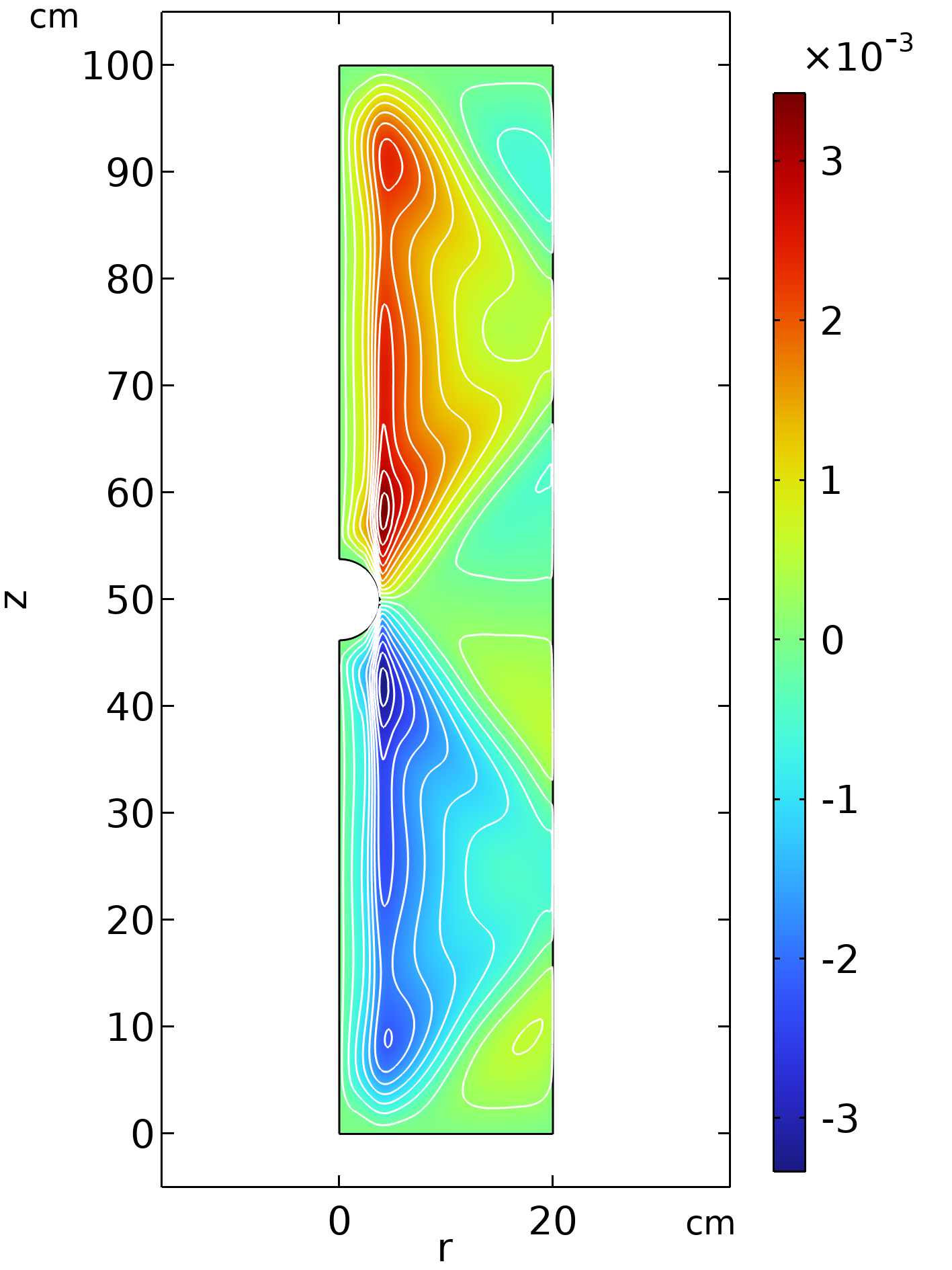} 
%        \caption{} \label{chiraltorque1}
    \end{subfigure}
%    \hfill
    \begin{subfigure}[t]{0.5\textwidth}
%       \centering
        \includegraphics[width=\linewidth]{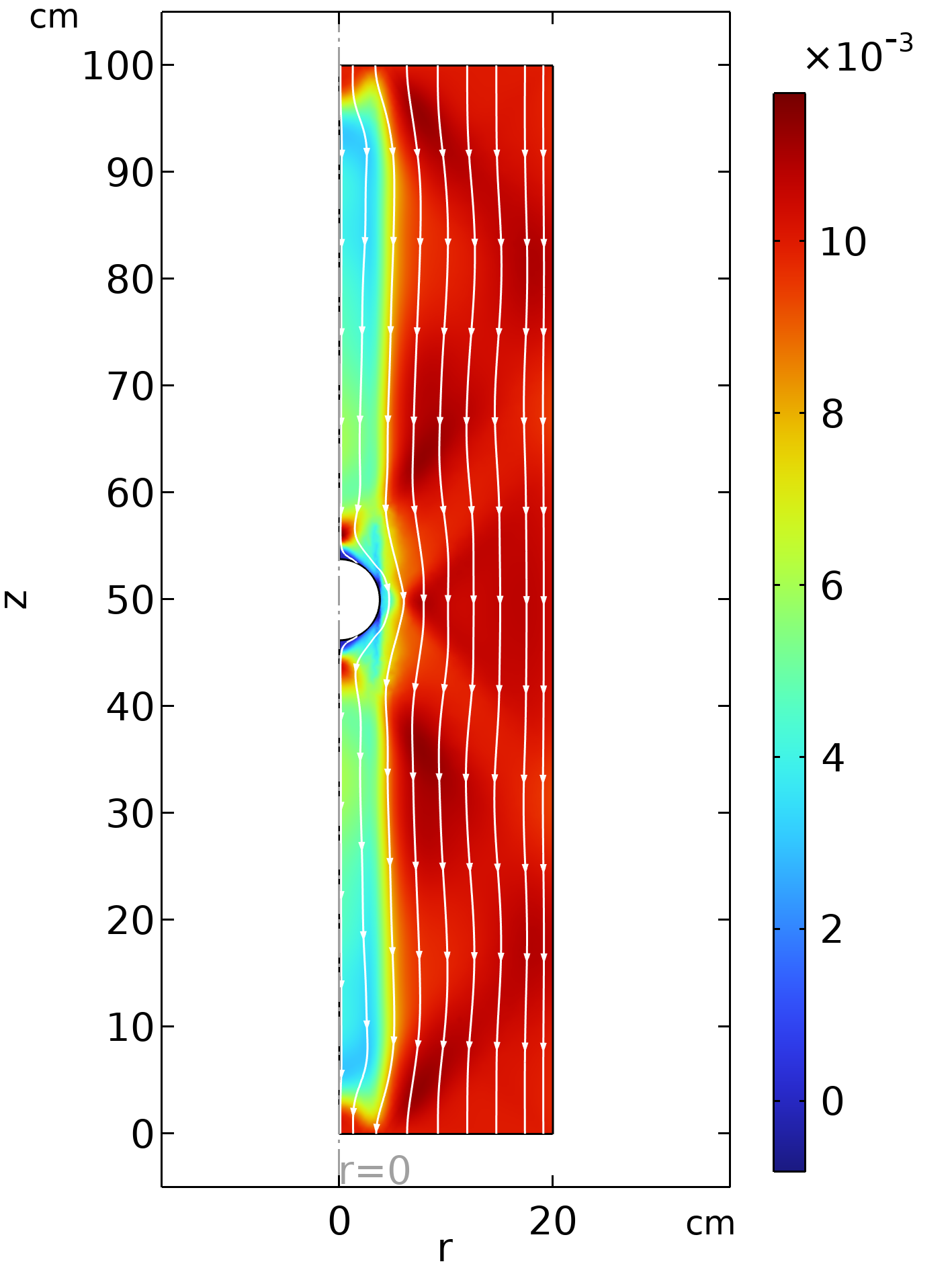} 
%        \caption{Spatial distribution of the observables $v_p, \sigma^{\textrm{ch}}$ and $v^{\textrm{ch}}$.} \label{vpschvchN}
    \end{subfigure}
     %  \vspace{1cm}
%    \begin{subfigure}[t]{\textwidth}
%    \centering
%        \includegraphics[width=\linewidth]{example-image-c.pdf} 
%        \caption{Price regulation} \label{fig:timing3}
%    \end{subfigure}
    \caption{\label{velocity2Dazimaxialeta4}{Distribution of azimuthal (left panel) and axial velocity (right panel) in an odd viscous
    liquid described by the constitutive law \rr{sigma4} ($\eta_o = 0$ and $\eta_4 = 1$ g/(cm sec)), moving slowly and meeting an immobile sphere (of radius $3.8$ cm) located at elevation $z=50$ cm 
 at the center axis of a cylinder. 
   Liquid enters from the top ($z=100$ cm) and exits at the bottom ($z=0$). The sphere is not allowed to rotate. 
   \textbf{Left:} Counter-rotation of liquid takes place above and below
    the sphere in the azimuthal direction. As observed in \citep[Fig.4(d)]{Khain2022} for the analogous problem  in Stokes flow, there are regions where the sense of rotation changes with distance from 
    the axis, i.e. the azimuthal velocity lobes that are additional to the ones of Fig. \ref{velocity2Dazimaxial}. 
    \textbf{Right:} A column whose generators are parallel to the cylinder axis and 
 circumscribes the sphere is also visible in the right panel. }
  } 
\end{figure}

\subsection{Axisymmetric inertial-like waves when $\eta_o \equiv 0$}
We express the constitutive law \rr{sigma4} in cylindrical coordinates
\be \label{sigma4rphi}
\bm{\sigma}' = \eta_4 
\left(\begin{array}{ccc}
0 & 0  & -(\frac{1}{r}\partial_\phi v_z + \partial_z v_\phi)\\
0 & 0  & \partial_r v_z + \partial_z v_r\\
 -(\frac{1}{r}\partial_\phi v_z + \partial_z v_\phi)&\partial_r v_z + \partial_z v_r &0
\end{array}
\right)
\ee
and repeat the construction of axial waves of section \ref{sec: inertial}.
The linearized equations of motion (see Appendix \ref{sec: appendix1}) become
\begin{align} \label{nsreta4}
-i\omega v_r & = -\frac{1}{\rho}\frac{\partial p'}{\partial r} + \nu_4 k^2 v_\phi
,  \\
-i\omega v_\phi & = 
-\nu_4\left[ \frac{1}{r} \frac{\partial}{\partial r} \left( r \frac{\partial v_r}{\partial r}\right)   - \frac{v_r}{r^2}\right] - \nu_4 k^2 v_r ,
\label{nsphieta4}
\\
-i\omega v_z& = -\frac{i k}{\rho} p' - ik \nu_4\frac{1}{r}\frac{\partial}{\partial r}(r v_\phi)
, \label{nszeta4}
\end{align}
where we simplified \rr{nsphieta4} by employing the incompressibility condition \rr{incrphi}. 
Introducing the linear operator \rr{L}
$
\mathcal{L} = \partial_r^2 + \frac{1}{r}\partial _r - \frac{1}{r^2}
$
the $r$ and $\phi$ momentum equations become
\begin{align} \label{systemL1eta4}
-i\omega v_r &= - i\frac{\omega}{k^2} \mathcal{L} v_r  +\nu_4(\mathcal{L} +k^2) v_\phi , \\
-i\omega v_\phi &= -\nu_4(\mathcal{L} +k^2) v_r. \label{systemL2eta4}
\end{align}
It is clear that the structure of these equations is identical to \rr{systemL1} and \rr{systemL2}
with the exception of the $k^2$ terms in the round brackets.  
The velocities $v_r$ and $v_\phi$ are again the Bessel functions, 
$v_r = AJ_1(\kappa r), v_\phi = BJ_1(\kappa r)$ (considering only the case where the origin is included
in the domain). 
System \rr{systemL1eta4} and \rr{systemL2eta4} has a solution when the determinant 
$-\kappa^{4} k^{2} \nu_4^{2}+\left(2 k^{4} \nu_4^{2}+\omega^{2}\right) \kappa^{2}-k^{2} \left(k^{2} \nu_4 -\omega \right) \left(k^{2} \nu_4 +\omega \right)$
of the coefficients 
of the resulting linear system
vanishes (for a derivation see Eq. \rr{ABeta40}). For real $\omega$ and $k$, $\kappa$ satisfies
\be \label{lambda2eta4}
\kappa^2 = \frac{\omega^{2}+2 k^{4} \nu_4^{2}\pm\omega  \sqrt{8 k^{4} \nu_4^{2}+\omega^{2}}}{2 k^{2} \nu_4^{2}}.
\ee 
There are two imaginary and two real roots when $\omega^2 - \nu_4^2k^4>0$.
There are four real $\kappa$ roots in \rr{lambda2eta4} when $\omega^2 - \nu_4^2k^4<0$
(the derivation of both is discussed below and summarized in Table \ref{table: roots}). 
This is to be expected from the form of the operator 
$\mathcal{S}$ in \rr{Leta4}
that determines the character of solutions. 
In constrast, in case \rr{lambda2} 
there are two real and two imaginary roots. Solving \rr{lambda2eta4} for $\omega$ one recovers 
\rr{dispeta4} with $k_x^2+k_y^2$ replaced by $\kappa^2$. 

\begin{figure}
\vspace{-5pt}
\begin{center}
\includegraphics[height=2.4in,width=4in]{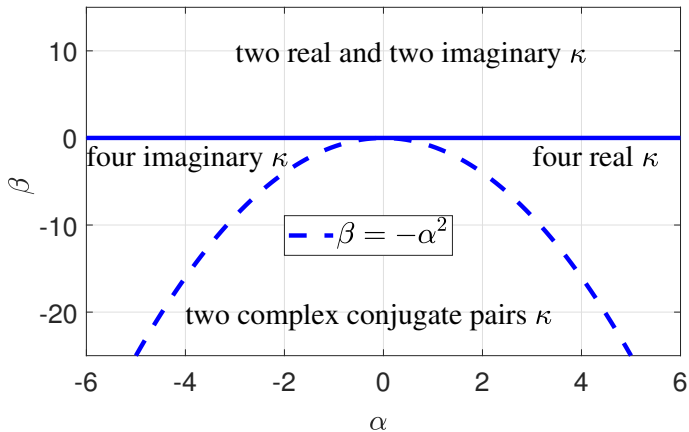}
\vspace{-0pt}
\end{center}
\caption{\label{roots}{Behavior of roots $\kappa$ of Eq. \rr{lambda2eta4eta0} \& \rr{lambda2eta4eta02} in the parameter space $(\alpha,\beta)$ defined in Eq. \rr{alphabeta}. 
This includes the oscillatory
Bessel functions for real $\kappa$, exponentially increasing/decreasing Bessel functions for $\kappa$ imaginary
and exponential-oscillating Bessel functions for complex $\kappa$.}}  
\vspace{-0pt}
\end{figure}

\subsection{Combined effect of $\eta_4$ and $\eta_o$ on axisymmetric inertial-like waves}
It turns out that a richer behavior is present when both $\eta_o$ and $\eta_4$
stress tensors \rr{sigma0} and \rr{sigma4rphi} are considered. Superposing the right hand sides 
of \rr{nsr}-\rr{nsz} to \rr{nsreta4}-\rr{nszeta4} and carrying-out the same manipulations as above
with the same linear operator \rr{L}
$
\mathcal{L} = \partial_r^2 + \frac{1}{r}\partial _r - \frac{1}{r^2},
$
the $r$ and $\phi$ momentum equations become
\begin{align} \label{systemL1eta4eta0}
-i\omega v_r &= - i\frac{\omega}{k^2} \mathcal{L} v_r  +\nu_4(\mathcal{L} +k^2) v_\phi-\nu_o\mathcal{L}v_\phi, \\
-i\omega v_\phi &= -\nu_4(\mathcal{L} +k^2) v_r + \nu_o\mathcal{L}v_r. \label{systemL2eta4eta0}
\end{align}
System \rr{systemL1eta4eta0} and \rr{systemL2eta4eta0} has a solution when the determinant 
$-\left(\nu_o -\nu_4 \right)^{2} k^{2} \kappa^{4}+\left(-2 k^{4} \nu_o \nu_4 +2 k^{4} \nu_4^{2}+\omega^{2}\right) \kappa^{2}-k^{2} \left(k^{2} \nu_4 -\omega \right) \left(k^{2} \nu_4 +\omega \right)$
of the coefficients 
of the resulting linear system
\be \label{ABeta40}
\frac{iBk^{4} \nu_4 +i \left(\nu_o -\nu_4 \right) \kappa^{2} B k^{2}-A \,\kappa^{2} \omega}{k^{2} \omega}-A =0, \quad \textrm{and} 
\quad -\frac{i A \left(\left(\nu_o -\nu_4 \right) \kappa^{2}+k^{2} \nu_4 \right)}{\omega}-B = 0
\ee
vanishes. 
The velocity field is again expressed with respect to 
Bessel functions $v_r = AJ_1(\kappa r)$ and $ v_\phi = BJ_1(\kappa r)$. 
For real $\omega$ and $k$ the eigenvalue $\kappa$ is 
\be \label{lambda2eta4eta0}
\kappa^2 = \frac{\pm\omega  \sqrt{4 \left(\nu_o -\nu_4 \right) \left(\nu_o -2 \nu_4 \right) k^{4}+\omega^{2}}+\omega^{2}-2 k^{4} \nu_4 \left(\nu_o -\nu_4 \right)}{2 \left(\nu_o -\nu_4 \right)^{2} k^{2}}.
\ee 
To understand the structure of solutions we define the frequency squared parameter 
$\alpha$ and quartic power of frequency $\beta$ of the form 
\be \label{alphabeta}
\alpha = \omega^{2}+2 k^{4} \nu_4 \left(\nu_4 -\nu_o \right), \quad 
\beta = 4k^4(\nu_o-\nu_4)^2(\omega^2 - \nu_4^2k^4). 
\ee
With this notation Eq. \rr{lambda2eta4eta0} simplifies to 
\be \label{lambda2eta4eta02}
2(\nu_o - \nu_4)^2 k^2 \kappa^2 = 
\alpha \pm \sqrt{\alpha^2 + \beta}.
\ee

In Fig. \ref{roots} we display all possible $\kappa$ behaviors inherent in \rr{lambda2eta4eta0}
and \rr{lambda2eta4eta02} and two cases are summarized in Table \ref{table: roots}. 
Thus, referring to Fig. \ref{roots}, when $\eta_4 \equiv 0$, $\beta$ is positive
and there are two real and two imaginary roots $\kappa$ (case considered in section \ref{sec: inertial}). When
$\eta_o \equiv 0$ (this is the case \rr{lambda2eta4}), $\alpha$ is always positive and $\alpha^2+\beta>0$. 
Thus, when $\beta<0$ there are four real $\kappa$ roots. In the opposite case two imaginary and two real. 
In the elliptic case $\eta_o = 2\eta_4$ there can be four imaginary or two imaginary and two real $\kappa$, 
as this case is tabulated in Table \ref{table: roots}. The fact that the operator 
$\mathcal{S}$ in \rr{Leta4}
is hyperbolic when $\nu_o=0$ and elliptic when $\nu_o = 2\nu_4$ is clearly reflected in the type of 
roots $\kappa$ and thus the form of the velocity field. 

Finally, of interest might also be the parabolic case where $\nu_o = \nu_4$. This case is separate
from \rr{lambda2eta4eta0} since the equations determining $\kappa$ are of second order in spatial 
derivatives (they are fourth order in the case \rr{lambda2eta4eta0}). We obtain 
\be
\kappa^2 = \frac{\left(k^{4} \nu_4^{2}-\omega^{2}\right) k^{2}}{\omega^{2}},
\ee
giving rise to either two real $\kappa$ or two imaginary $\kappa$. The dispersion relation reads
\be
\omega = \pm \frac{\nu_4 k^3}{\sqrt{\kappa^2 + k^2}}. 
\ee

\begin{table}
  \begin{center}
\def~{\hphantom{0}}
  \begin{tabular}{c|cc|c|c}
%  \multicolumn{2}{l}{} & \multicolumn{2}{c}{$-3$ \hfil 2 \hfil } \\\hline
&&$\omega^2<\nu_4^2k^4$&$\nu_4^2k^4<\omega^2<2\nu_4^2k^4$&$\omega^2>2\nu_4^2k^4$\\\hline
& $\alpha$ & $+$ & $+$&$+$ \\
$\eta_o= 0$ & $\beta$ & $-$ & $+$&$+$ \\
&$\kappa$ & 4 Re &2 Im+2 Re & 2 Im + 2 Re \\\hline
& $\alpha$ & $-$ & $-$&$+$ \\
$\eta_o=2\eta_4$ & $\beta$ & $-$ & $+$&$+$ \\
&$\kappa$ & 4 Im &2 Im+2 Re & 2 Im + 2 Re 
%& $\alpha$ & $+$ & $+$&$+$ \\
%$\eta_4= 0$ & $\beta$ & $+$ & $+$&$+$ \\
%&$\kappa$ & 2Im+2 Re &2 Im+2 Re & 2 Im + 2 Re \\\hline
  \end{tabular}
\caption{Types of roots $\kappa$ from Eq. \rr{lambda2eta4eta0}/\rr{lambda2eta4eta02} where $\kappa^2 \propto 
\alpha \pm \sqrt{\alpha^2 + \beta}$ according to the sign
of the parameters $\alpha $ and $\beta$ defined in Eq. \rr{alphabeta} for two special choices of the 
odd viscosity parameters. In all cases $\alpha^2 + \beta>0$. Re = Real $\kappa$. Im = Imaginary $\kappa$.}
  \label{table: roots}
  \end{center}
\end{table}

\section{\label{sec: comparison}Comparison to the recent literature of odd viscous liquids}
We conclude this paper by summarizing a few results, taken from the recent literature, that 
are related to the present paper, where some of them also appeared in the foregoing analysis. 

Consider the joint $\eta_o$ and $\eta_4$ effects. The identification $\eta_o = 2\eta_4$ has been employed in the literature as a means to incorporate
effects of both odd coefficients $\eta_o$ and $\eta_4$ and to simplify the presentation \citep{Khain2022}. 
The dispersion relation \rr{dispeta4}
$\omega = -k^{2}\left(\left(\nu_o -2 \nu_4 \right) \cos^{2}\theta - \nu_o + \nu_4 \right) \cos  \theta  $ simplifies significantly giving
\be
\omega = \nu_4 k k_z,
\ee
which recovers the dispersion relation of a Hamiltonian formulation of 
spinning molecules \citep[Eq. (8)]{Markovich2021}. 

As discussed above, the counter-rotation of an odd viscous liquid below and above 
a slowly-moving sphere in Fig. \ref{velocity2Dazimaxial} ($\eta_o\neq 0, \eta_4=0$) agrees with the observations of \citep[Fig.4(c)]{Khain2022} for the analogous problem in Stokes flow. Similarly,
the presence of additional lobes of azimuthal velocity in Fig. \ref{velocity2Dazimaxial}
($\eta_o =0, \eta_4\neq0$), was also observed in Stokes flow \citep[Fig.4(d)]{Khain2022}. 

In general, large values of the odd viscosity lead to recirculating patterns, see \citep[Fig. 6(h)]{Khain2022}
in agreement with the restoring effect of an odd viscous liquid discussed in the introduction of the 
present paper. The combination $\eta_o = 2\eta_4$ leads
to swirling flow with a large azimuthal velocity component. This type of flow is stabilizing. Thus, 
we can say for instance that a cloud sedimenting in an odd viscous liquid \citep[Fig. 6(b)]{Khain2022}
is circumscribed by its own Taylor column and it does not disintegrate.

\section{Discussion}
In this article we showed that inertial-like waves and Taylor columns are generated in a three-dimensional odd viscous liquid. Both odd coefficients $\eta_o$ and $\eta_4$ that appear in the constitutive laws
\rr{sigma1} and \rr{sigma4} are nicely tucked-away in a compact differential operator $\mathcal{S}$
Eq. \rr{Leta4} and result in the odd form of the Navier-Stokes equations \rr{DvDt2} and \rr{dtcurlv2}. 
The flow that arises from the consideration of plane-polarized waves is a Beltrami flow. Thus, its
helicity is conserved. The flow field determined by 
three dimensional axisymmetric waves in an odd $\eta_o$ and $\eta_4$ liquid has a structure that 
is determined by the eigenvalues $\kappa$ in the argument of the Bessel function $J_1(\kappa r)$ (or $Y_1(\kappa r)$) and a classification of different behaviors is displayed in 
Fig. \ref{roots}. Thus, the velocity field can be oscillatory, exponential increasing/decaying or their combination
thereof. }

Considering the behavior of liquids with a single $\eta_o$ odd viscosity coefficient, we can isolate two
important results that were developed here. 
First, we observe inertial oscillations downstream a slowly moving body whose
theoretically determined wavelength Eq. \rr{wavelength} is in agreement with 
its estimate from solutions of the full Navier-Stokes equations (in an analogous manner to the experiments
of \citet{Long1953} which were in agreement with the theory of inertial oscillations in rigidly rotating liquids, 
cf. \cite[plate 24]{Batchelor1967}). Second, we observe Taylor-column-like behavior when a sphere
slowly moves along the axis of anisotropy, and this also resembles the behavior of a rigidly-rotating liquids, 
for instance, in the experiments of 
\citet{Maxworthy1970}. The latter behavior is to be expected since, when the wavevector is 
perpendicular to the anisotropy axis, the flow becomes effectively two-dimensional in agreement 
with the (modified) Taylor-Proudman theorem we developed here, suitable for odd viscous liquids. 
At the same time helicity segregation signals the generation of inertial-like waves at the interior of the column
where information is transported along the anisotropy axis, above the body and below the body at the 
group velocity.  

{Our theoretical discussion was predominantly centered on an odd viscous liquid with zero shear viscosity
(in the numerical simulations of the Navier-Stokes equations shear viscosity was however, small but non-zero).
Shear viscosity would just endow the frequency $\omega$ with an imaginary part, as is the case 
in rigidly-rotating liquids, see for instance \citep[p. 86, Eq. (72)]{Chandrasekhar1961}.}

There is a large number of unexplored phenomena in three-dimensional odd viscous liquids 
associated with the findings in this paper. To name a few, what is the effect of shear viscosity on Taylor columns, 
and thus the establishment of Ekman and Stewartson layers. The effect of different flow conditions
(eg. flow incident on an finite length obstacle also generating Taylor columns), the effect of different
material properties (eg. a liquid droplet instead of a solid sphere 
rising slowly in such a liquid, cf. \citep{Bush1994, Bush1995}), a freely suspended sphere rising slowly in such a liquid and finally an experimental realization of these effects. Also, determination of criteria for Taylor-column formation in bounded and unbounded domains. To understand the diversity behind these issues in the case of rotating liquids, one can be apprised by the review of \cite{Bush1994}. 

The formulation developed in this paper can prove to be useful in many areas 
of research. For instance, one could envision the following application of
odd viscous liquid Taylor columns to materials science: 
Semiconducting nanowires, employed in diverse fields
such as biological molecule sensing and living cell probing, 
are manufactured by the vapor-liquid-solid growth technique, 
whereby a liquid-alloy droplet increases in size by absorbing
material from a vapor phase. Hydrodynamics has been shown to be an important factor in this process \citep{schwalbach2012}. 
Disruption of growth may take place in the form of instabilities leading the nanowire to adopt distorted
shapes or to develop random extrusions. These instability mechanisms 
may however become suppressed in the presence of odd viscosity. 
Slow impinging flow on a nanowire will circumscribe each one of them to its own
Taylor column, thus providing a unidirectional guide for droplet-alloy growth,
circumventing the generation of instabilities.   
Thus, a fundamental related open question concerns the effect odd viscosity has on convection.

\noindent
\textbf{Acknowledgments}\\ 
This research was supported by the Center for Bio-Inspired Energy Science, an Energy Frontier Research Center funded by the US Department of Energy, Office of Science, Basic Energy Sciences under Award No. DE-SC0000989. We thank Leticia Lopez and four anonymous referees for comments that improved the manuscript. 
\\\\
\textbf{Declaration of Interests}\\ The authors report no conflict of interest.

\appendix

\section{\label{sec: oddstress}Odd viscous stress}
In two dimensions it has been shown \citep{Kirkinis2022a} that the odd viscous stress tensor and thus the total stress, can succinctly be written 
in the form
\be \label{seso}
\bm{\sigma} =-p\bm{I}+
%=  \left( \begin{array}{cc}
% 1 & - \frac{\eta^o}{\eta^e} \\
%  \frac{\eta^o}{\eta^e} & 1
% \end{array} \right) \sigma^e
2 \left( \begin{array}{cc}
 \eta_e & -\eta_o \\
 \eta_o & \eta_e
 \end{array} \right)
{D}, 
%\equiv
% \left( \begin{array}{cc}
% \eta^e & -\eta^o \\
% \eta^o & \eta^e
% \end{array} \right)
%  \left( \begin{array}{cc}
% 2u_x & u_z+w_x \\
% u_z+w_x & 2w_z
% \end{array} \right)
\ee
in terms of the two-dimensional rate-of-strain tensor 
\be 
{D}_{ij} = \frac{1}{2}\left(\frac{\partial u_i}{\partial x_j} + \frac{\partial u_j}{\partial x_i}\right), \quad i, j=1,2
\ee
for an incompressible liquid. 
Expression \rr{seso} is important because it clarifies the origin of physical effects appearing on a free surface
of an odd viscous liquid. 
In this section we would like to obtain an analogous expression for the three dimensional odd viscous
stress \rr{sigma1} and \rr{sigma4}. To this end, 
define the three dimensional rotation matrix
\be
R = 
\left(\begin{array}{ccc}
0 & -1 & 0\\
1 & 0  & 0\\
0&0&0
\end{array}
\right)
\ee
so that $R^T = -R$. A little work (using the notation 
$
\sigma^o = 2\eta_o 
\left(\begin{array}{ccc}
-D_{12} & \frac{D_{11} - D_{22}}{2} & 0\\
\frac{D_{11} - D_{22}}{2} &D_{12}  & 0\\
0&0&0
\end{array}
\right)
$
etc. here $D_{ij}$ should be understood as the upper left two by two block of the three-dimensional rate-of-strain tensor)
shows that 
\be \label{RDDR}
\sigma^o = {\eta_o} \left[ RD + (RD)^T\right] = {\eta_o} \left[ RD - DR\right]. 
\ee
We can repeat the above for the tensor \rr{sigma4} (here $D$ should be understood as the rate-of-strain tensor without the upper left two by two block) and obtain 
\be \label{RDDR4}
\sigma^4 = {\eta_4} \left[ RD + (RD)^T\right] = {\eta_4} \left[ RD - DR\right]. 
\ee
When $\eta_o = \eta_4$ (this is the parabolic case for the operator $\mathcal{S}$ of section \ref{sec: eta4})
we can employ a compact notation for the stress tensor incorporating both odd coefficients and the shear 
viscosity $\eta_e$ by splitting the even stress tensor 
$\sigma^e = 2\eta_e D = \eta_e(ID + DI)$, where $I$ is the unit matrix to obtain
\be
\sigma =-pI+ (\eta_eI + \eta_o R) D + D (\eta_eI - \eta_o R), 
\ee
where ${D}$ is the three-dimensional rate-of-strain tensor
\be
{D}_{ij} =\frac{1}{2}\left( \frac{\partial u_i}{\partial x_j} + \frac{\partial u_j}{\partial x_i}\right), \quad i, j= 1,2,3.
\ee
Alternatively, define the viscosity matrix 
\be
\hat{\eta} = 
\left(\begin{array}{ccc}
\eta_e & -\eta_o & 0\\
\eta_o & \eta_e  & 0\\
0&0&\eta_e
\end{array}
\right)
\ee
Then, the full stress tensor in three dimensions obtains the form
\be \label{sigma3D}
\sigma =-pI+ \hat{\eta} D + D\hat{\eta}^T, 
\ee
which is the analogue of \rr{seso}. 
In two dimensions $D_{11}= - D_{22}$ and Eq. \rr{sigma3D} immediately reduces to Eq. \rr{seso}. 

One immediate consequence of writing the odd stress tensor in the form \rr{RDDR} or \rr{RDDR4} is that
it makes clear the effect of odd viscosity on dissipation of energy. The kinetic energy density 
\cite[\S 16]{Landau1987} 
$$
\dot{\textrm{E}}_{\textrm{kin}} = - \textrm{tr}\left({\sigma}^o D\right)
$$
vanishes immediately since ${\sigma}^o$ in \rr{RDDR} is the product of a symmetric and an antisymmetric matrix
(tr denotes the trace of a matrix), and likewise for $\sigma^4$.

\section{\label{sec: appendix1}Linearized equations of motion}
\begin{align}
\partial_t v_r &= -\frac{1}{\rho}\frac{\partial p'}{\partial r}  +\frac{1}{\rho}\left[ \frac{1}{r}\partial_r (r\sigma_{rr}) + \frac{1}{r}\partial_\phi \sigma_{r\phi} +\partial_z \sigma_{rz} - 
\frac{1}{r}\sigma_{\phi\phi}                                    \right] ,  \\
\partial_t v_\phi &= -\frac{1}{\rho r}\frac{\partial p'}{\partial \phi}  +\frac{1}{\rho}
\left[ \frac{1}{r^2}\partial_r (r^2\sigma_{\phi r}) + \frac{1}{r}\partial_\phi \sigma_{\phi \phi} +\partial_z \sigma_{\phi z} + 
\frac{1}{r}(\sigma_{r\phi} - \sigma_{\phi r})                                    \right], \\
\partial_t v_z &= -\frac{1}{\rho}\frac{\partial p'}{\partial z}  +\frac{1}{\rho}\left[ \frac{1}{r}\partial_r (r\sigma_{zr}) + \frac{1}{r}\partial_\phi \sigma_{z\phi} +\partial_z \sigma_{zz}                               \right] .
\end{align}
Note that the definition of the stress tensor in fluid mechanics (cf. \citep{Landau1987}) 
differs from its definition in the continuum mechanics literature where it is defined as the transpose. Here we follow
the fluid mechanics notation, as this arises for instance in \citep{Landau1987}.

\section{\label{sec: rotating}Basic facts about rotating fluids}
The Rossby and Ekman numbers are
\be \label{RoEk}
Ro = \frac{U}{\Omega L}, \quad E = \frac{\nu}{\Omega L}.
\ee
In the axisymmetric case where $\partial_\phi =0$, the Taylor-Proudman theorem for the 
geostrophic equations
\be
-\Omega v_\phi = - \frac{1}{\rho} \frac{\partial p}{\partial r}, \quad  \Omega v_r=0, \quad 0= \frac{\partial p}{\partial r}
\ee
implies that $v_r\equiv 0$ and thus the streamlines are spirals that wound around circular cylinders
\citep{Yih1959}. 

\subsection{Taylor columns}
The geostrophic equations, in Cartesian coordinates, written in the form
\be \label{TP0}
\frac{1}{\rho}\frac{\partial p}{\partial x} = -\Omega v, \quad \frac{1}{\rho}\frac{\partial p}{\partial y} = \Omega u, \quad \frac{\partial p}{\partial z} = 0,  
\ee
show that the pressure $p$ is a streamfunction and thus constant on a streamline of the flow.
A finite-length cylinder with generators parallel to the rotating axis and moving horizontally in a rotating liquid will thus be accompanied by a liquid velocity parallel to its 
generators and a column will accompany its motion \cite[\S 12.2]{Yih1988}.
Inside the column the velocity can be zero, although viscous liquids are accompanied with 
special flows where the velocity does not vanish \citep{Moore1968}.

Separate two-dimensional flows exist inside and outside the Taylor column. Liquid cannot be transferred
between these two regions. This is clear in the axisymmetric case where $v_r$ vanishes everywhere (outside the Taylor column). Experimentally dye that is outside the Taylor column cannot enter
and dye inside does not exit \cite[Fig.16.2]{Tritton1988}. The flow inside the Taylor column is determined 
by taking into account the thin shear layers that develop on the lateral surface of the Taylor column and 
the Ekman boundary-layers on the body and the boundaries \citep{Moore1968}.

\subsection{Elasticity induced by rotation of an inviscid liquid}
Consider a particle of unit mass that moves with a speed $v$ perpendicular to the axis of rotation. 
Momentum conservation gives 
\be
\frac{v^2}{r} = 2\Omega v.
\ee
Solving for $r$ we obtain $r = \frac{v}{2\Omega}$. This implies that the locus of the particle
is a circle. It goes around the circle twice during every revolution of the liquid with period $T = 2\pi r/ v = \pi/\Omega$ (the vorticity of the liquid in rigid-body rotation is twice the angular velocity of rotation)
\cite[\S16.6]{Tritton1988}. 

The effect of this constraining tendency is to support inertial waves. As remarked above, inertial waves
exist only when $\omega < 2\Omega$ since $\kappa = k\sqrt{\frac{4\Omega^2}{\omega^2} -1}$
must be real. 

The relations (Taylor-Proudman theorem)
\be
\frac{\partial u}{\partial z} = \frac{\partial v}{\partial z} =0, 
\ee
do not allow vortex twisting (the liquid velocity approaching a finite obstacle does not
change relative to the obstacle - vorticity is not generated), see \cite[Fig.16.5]{Tritton1988}. 
This means that the background vorticity resists twisting. On the other hand the condition
\be
\frac{\partial w}{\partial z} =0, 
\ee
resists vortex stretching (vortex tubes do not thin to increase vorticity see \cite[Fig.16.6]{Tritton1988}). 

Another view of the same effect is to displace a circular ring of fluid outward to a new position $r$. 
The circulation $2\pi v r$ along that ring remains the same according to Kelvin's theorem. Thus, the ring's
$v$ and $v^2/r$ will be smaller to the new position. However the liquid $v^2/r$ was larger there
and it was balanced exactly by a pressure gradient, which, now sees the lower $v^2/r$ of the ring. 
Thus, it will push the ring back towards its original position. The pressure at this position will push
it again outwards and the ring will experience an oscillatory motion \cite[\S 5]{Yih1988}. See
also \citep[Chapter 1]{Davidson2013book} for a more detailed explanation of the same effect. 
A similar discussion about the restoring effect of the Coriolis force can be traced back to \citep[\S 7.6]{Batchelor1967}.

\def\cprime{$'$}


\begin{thebibliography}{44}
\expandafter\ifx\csname natexlab\endcsname\relax\def\natexlab#1{#1}\fi
\def\au#1{#1} \def\ed#1{#1} \def\yr#1{#1}\def\at#1{#1}\def\jt#1{\textit{#1}}
  \def\bt#1{#1}\def\bvol#1{\textbf{#1}} \def\vol#1{#1} \def\pg#1{#1}
  \def\publ#1{#1}\def\arxiv#1{#1}\def\org#1{#1}\def\st#1{\textit{#1}}

\bibitem[Abanov {\em et~al.\/}(2018)Abanov, Can \& Ganeshan]{Abanov2018}
{\sc \au{Abanov, A.}, \au{Can, T.} \& \au{Ganeshan, S.}} \yr{2018}  \at{Odd
  surface waves in two-dimensional incompressible fluids}.  \jt{SciPost
  Physics}  \bvol{5}~(1),  \pg{010}.

\bibitem[Asselin \& Young(2020)]{Asselin2020}
{\sc \au{Asselin, O.} \& \au{Young, W.R.}} \yr{2020}  \at{Penetration of
  wind-generated near-inertial waves into a turbulent ocean}.  \jt{{Journal of
  Physical Oceanography}}  \bvol{50}~(6),  \pg{1699--1716}.

\bibitem[Avron(1998)]{Avron1998}
{\sc \au{Avron, J.E.}} \yr{1998}  \at{Odd viscosity}.  \jt{{Journal of
  Statistical Physics}}  \bvol{92}~(3-4),  \pg{543--557}.

\bibitem[Avron {\em et~al.\/}(1995)Avron, Seiler \& Zograf]{Avron1995}
{\sc \au{Avron, J.E.}, \au{Seiler, R.} \& \au{Zograf, P.G.}} \yr{1995}
  \at{{Viscosity of quantum Hall fluids}}.  \jt{{Physical Review Letters}}
  \bvol{75}~(4),  \pg{697}.

\bibitem[Banerjee {\em et~al.\/}(2017)Banerjee, Souslov, Abanov \&
  Vitelli]{Banerjee2017}
{\sc \au{Banerjee, D.}, \au{Souslov, A.}, \au{Abanov, A.G.} \& \au{Vitelli,
  V.}} \yr{2017}  \at{Odd viscosity in chiral active fluids}.  \jt{{Nature
  Communications}}  \bvol{8}~(1),  \pg{1573}.

\bibitem[Batchelor(1967)]{Batchelor1967}
{\sc \au{Batchelor, G.~K.}} \yr{1967} {\em An introduction to fluid
  dynamics\/}.  \publ{Cambridge: Cambridge University Press}.

\bibitem[Bire {\em et~al.\/}(2022)Bire, Kang, Ramadhan, Campin \&
  Marshall]{Bire2022}
{\sc \au{Bire, S.}, \au{Kang, W.}, \au{Ramadhan, A.}, \au{Campin, J.-M.} \&
  \au{Marshall, J.}} \yr{2022}  \at{Exploring ocean circulation on icy moons
  heated from below}.  \jt{Journal of Geophysical Research: Planets,}
  \bvol{127},  \pg{e2021JE007025}.

\bibitem[Bush {\em et~al.\/}(1995)Bush, Stone \& Bloxham]{Bush1995}
{\sc \au{Bush, J.W.M.}, \au{Stone, H.A.} \& \au{Bloxham, J.}} \yr{1995}
  \at{Axial drop motion in rotating fluids}.  \jt{{Journal of Fluid Mechanics}}
   \bvol{282},  \pg{247--278}.

\bibitem[Bush {\em et~al.\/}(1994)Bush, Stone \& Tanzosh]{Bush1994}
{\sc \au{Bush, J.W.M.}, \au{Stone, H.A.} \& \au{Tanzosh, J.P.}} \yr{1994}
  \at{Particle motion in rotating viscous fluids: Historical survey and recent
  developments}.  \jt{Current Topics in The Physics of Fluids}  \bvol{1},
  \pg{337--355}.

\bibitem[Chandrasekhar(1961)]{Chandrasekhar1961}
{\sc \au{Chandrasekhar, S.}} \yr{1961} {\em Hydrodynamic and hydromagnetic
  stability\/}.  \publ{Oxford University Press}.

\bibitem[Dahler \& Scriven(1961)]{Dahler1961}
{\sc \au{Dahler, J.S.} \& \au{Scriven, L.E.}} \yr{1961}  \at{Angular momentum
  of continua}.  \jt{Nature}  \bvol{192},  \pg{36--37}.

\bibitem[Davidson(2013)]{Davidson2013book}
{\sc \au{Davidson, P.A.}} \yr{2013} {\em Turbulence in rotating, stratified and
  electrically conducting fluids\/}.  \publ{{Cambridge University Press,
  Cambridge}}.

\bibitem[Davidson(2014)]{Davidson2014}
{\sc \au{Davidson, P.A.}} \yr{2014}  \at{The dynamics and scaling laws of
  planetary dynamos driven by inertial waves}.  \jt{{Geophysical Journal
  International}}  \bvol{198}~(3),  \pg{1832--1847}.

\bibitem[Davidson \& Ranjan(2018)]{Davidson2018}
{\sc \au{Davidson, P.A.} \& \au{Ranjan, A.}} \yr{2018}  \at{On the spatial
  segregation of helicity by inertial waves in dynamo simulations and planetary
  cores}.  \jt{{Journal of Fluid Mechanics}}  \bvol{851},  \pg{268--287}.

\bibitem[Dransfeld {\em et~al.\/}(2009)Dransfeld, Dwane \& Zuur]{Dransfeld2009}
{\sc \au{Dransfeld, L.}, \au{Dwane, O.} \& \au{Zuur, A.F.}} \yr{2009}
  \at{{Distribution patterns of ichthyoplankton communities in different
  ecosystems of the Northeast Atlantic}}.  \jt{{Fisheries Oceanography}}
  \bvol{18}~(6),  \pg{470--475}.

\bibitem[Fruchart {\em et~al.\/}(2023)Fruchart, Scheibner \&
  Vitelli]{Fruchart2023}
{\sc \au{Fruchart, M.}, \au{Scheibner, C.} \& \au{Vitelli, V.}} \yr{2023}
  \at{Odd viscosity and odd elasticity}.  \jt{{Annual Review of Condensed
  Matter Physics}}  \bvol{14},  \pg{471--510}.

\bibitem[Ganeshan \& Abanov(2017)]{Ganeshan2017}
{\sc \au{Ganeshan, S.} \& \au{Abanov, A.G.}} \yr{2017}  \at{Odd viscosity in
  two-dimensional incompressible fluids}.  \jt{Physical Review Fluids}
  \bvol{2}~(9),  \pg{094101}.

\bibitem[Gao {\em et~al.\/}(2020)Gao, Chew, Marxen {\em et~al.\/}]{Gao2020}
{\sc \au{Gao, F.}, \au{Chew, J.W.}, \au{Marxen, O.} \& \au{others}} \yr{2020}
  \at{Inertial waves in turbine rim seal flows}.  \jt{{Physical Review Fluids}}
   \bvol{5}~(2),  \pg{024802}.

\bibitem[Greenspan(1968)]{Greenspan1968}
{\sc \au{Greenspan, H.P.}} \yr{1968} {\em The theory of rotating fluids\/}.
  \publ{Cambridge Univ Press}.

\bibitem[Khain {\em et~al.\/}(2022)Khain, Scheibner, Fruchart \&
  Vitelli]{Khain2022}
{\sc \au{Khain, T.}, \au{Scheibner, C.}, \au{Fruchart, M.} \& \au{Vitelli, V.}}
  \yr{2022}  \at{Stokes flows in three-dimensional fluids with odd and
  parity-violating viscosities}.  \jt{{Journal of Fluid Mechanics}}
  \bvol{934},  \pg{A23}.

\bibitem[Kirkinis(2017)]{Kirkinis2017}
{\sc \au{Kirkinis, E.}} \yr{2017}  \at{Magnetic torque-induced suppression of
  van-der-{W}aals-driven thin liquid film rupture}.  \jt{Journal of Fluid
  Mechanics}  \bvol{813},  \pg{991--1006}.

\bibitem[Kirkinis(2023)]{Kirkinis2023null}
{\sc \au{Kirkinis, E.}} \yr{2023}  \at{Null-divergence nature of the odd
  viscous stress for an incompressible liquid}.  \jt{{Physical Review Fluids}}
  \bvol{8}~(1),  \pg{014104}.

\bibitem[Kirkinis \& Andreev(2019)]{Kirkinis2019b}
{\sc \au{Kirkinis, E.} \& \au{Andreev, A.V.}} \yr{2019}  \at{{Odd
  viscosity-induced stabilization of viscous thin liquid films}}.  \jt{Journal
  of Fluid Mechanics}  \bvol{878},  \pg{169--189}.

\bibitem[Kirkinis {\em et~al.\/}(2022)Kirkinis, Mason \& Olvera de~la
  Cruz]{Kirkinis2022a}
{\sc \au{Kirkinis, E.}, \au{Mason, J.} \& \au{Olvera de~la Cruz, M.}} \yr{2022}
   \at{{Odd viscosity-induced passivation of Moffatt vortices}}.  \jt{Journal
  of Fluid Mechanics}  \bvol{950},  \pg{A19}.

\bibitem[Knobloch(2022)]{Knobloch2022}
{\sc \au{Knobloch, E.}} \yr{2022}  \at{Geostrophic turbulence and the formation
  of large scale structure}.  \bt{In {\em {Mathematical and Computational
  Models of Flows and Waves in Geophysics}\/}},  \pg{pp. 1--34}.
  \publ{Springer}.

\bibitem[Landau \& Lifshitz(1987)]{Landau1987}
{\sc \au{Landau, L.~D.} \& \au{Lifshitz, E.~M.}} \yr{1987} {\em Fluid
  Mechanics. {C}ourse of {T}heoretical {P}hysics, {V}ol. 6\/}.  \publ{Pergamon
  Press Ltd., London-Paris}.

\bibitem[Lifshitz \& Pitaevskii(1981)]{Landau1981}
{\sc \au{Lifshitz, E.~M.} \& \au{Pitaevskii, L.~P.}} \yr{1981} {\em {Course of
  theoretical physics. {V}ol. 10: Physical Kinetics}\/}.  \publ{Pergamon
  Press}.

\bibitem[Long(1953)]{Long1953}
{\sc \au{Long, R.R.}} \yr{1953}  \at{Steady motion around a symmetrical
  obstacle moving along the axis of a rotating liquid}.  \jt{{Journal of
  Atmospheric Sciences}}  \bvol{10}~(3),  \pg{197--203}.

\bibitem[Markovich \& Lubensky(2021)]{Markovich2021}
{\sc \au{Markovich, T.} \& \au{Lubensky, T.C.}} \yr{2021}  \at{Odd viscosity in
  active matter: microscopic origin and 3d effects}.  \jt{{Physical Review
  Letters}}  \bvol{127}~(4),  \pg{048001}.

\bibitem[Maxworthy(1970)]{Maxworthy1970}
{\sc \au{Maxworthy, T.}} \yr{1970}  \at{The flow created by a sphere moving
  along the axis of a rotating, slightly-viscous fluid}.  \jt{{Journal of Fluid
  Mechanics}}  \bvol{40}~(3),  \pg{453--479}.

\bibitem[Moffatt(1969)]{Moffatt1969}
{\sc \au{Moffatt, H.K.}} \yr{1969}  \at{The degree of knottedness of tangled
  vortex lines}.  \jt{{Journal of Fluid Mechanics}}  \bvol{35}~(1),
  \pg{117--129}.

\bibitem[Moffatt(1970)]{Moffatt1970}
{\sc \au{Moffatt, H.K.}} \yr{1970}  \at{Dynamo action associated with random
  inertial waves in a rotating conducting fluid}.  \jt{{Journal of Fluid
  Mechanics}}  \bvol{44}~(4),  \pg{705--719}.

\bibitem[Moore \& Saffman(1968)]{Moore1968}
{\sc \au{Moore, D.W.} \& \au{Saffman, P.G.}} \yr{1968}  \at{The rise of a body
  through a rotating fluid in a container of finite length}.  \jt{{Journal of
  Fluid Mechanics}}  \bvol{31}~(4),  \pg{635--642}.

\bibitem[Ogilvie(2013)]{Ogilvie2013}
{\sc \au{Ogilvie, G.I.}} \yr{2013}  \at{Tides in rotating barotropic fluid
  bodies: the contribution of inertial waves and the role of internal
  structure}.  \jt{{Monthly Notices of the Royal Astronomical Society}}
  \bvol{429}~(1),  \pg{613--632}.

\bibitem[Rinaldi(2002)]{Rinaldi2002Thesis}
{\sc \au{Rinaldi, C.}} \yr{2002}  \at{Continuum modeling of polarizable
  systems}. PhD thesis, Massachusetts Institute of Technology.

\bibitem[Schwalbach {\em et~al.\/}(2012)Schwalbach, Davis, Voorhees, Warren \&
  Wheeler]{schwalbach2012}
{\sc \au{Schwalbach, E.J.}, \au{Davis, S.H.}, \au{Voorhees, P.W.}, \au{Warren,
  J.A.} \& \au{Wheeler, D.}} \yr{2012}  \at{Stability and topological
  transformations of liquid droplets on vapor-liquid-solid nanowires}.
  \jt{{Journal of Applied Physics}}  \bvol{111}~(2),  \pg{024302}.

\bibitem[Soni {\em et~al.\/}(2019)Soni, Bililign, Magkiriadou, Sacanna,
  Bartolo, Shelley \& Irvine]{Soni2019}
{\sc \au{Soni, V.}, \au{Bililign, E.S.}, \au{Magkiriadou, S.}, \au{Sacanna,
  S.}, \au{Bartolo, D.}, \au{Shelley, M.J.} \& \au{Irvine, W.T.M.}} \yr{2019}
  \at{The odd free surface flows of a colloidal chiral fluid}.  \jt{Nature
  Physics}  \bvol{15}~(11),  \pg{1188--1194}.

\bibitem[Souslov {\em et~al.\/}(2019)Souslov, Dasbiswas, Fruchart,
  Vaikuntanathan \& Vitelli]{Souslov2019}
{\sc \au{Souslov, A.}, \au{Dasbiswas, K.}, \au{Fruchart, M.},
  \au{Vaikuntanathan, S.} \& \au{Vitelli, V.}} \yr{2019}  \at{Topological waves
  in fluids with odd viscosity}.  \jt{{Physical Review Letters}}
  \bvol{122}~(12),  \pg{128001}.

\bibitem[Tanzosh \& Stone(1994)]{Tanzosh1994}
{\sc \au{Tanzosh, J.P.} \& \au{Stone, H.A.}} \yr{1994}  \at{Motion of a rigid
  particle in a rotating viscous flow: an integral equation approach}.
  \jt{{Journal of Fluid Mechanics}}  \bvol{275},  \pg{225--256}.

\bibitem[Tritton(1988)]{Tritton1988}
{\sc \au{Tritton, D.J.}} \yr{1988} {\em Physical Fluid Dynamics\/}.
  \publ{Oxford University Press: Oxford}.

\bibitem[Truesdell \& Noll(1992)]{Truesdell1992}
{\sc \au{Truesdell, C.} \& \au{Noll, W.}} \yr{1992} {\em The nonlinear field
  theories of mechanics\/}, 2nd edn.  \publ{Berlin: Springer-Verlag}.

\bibitem[Whitham(1974)]{Whitham1974}
{\sc \au{Whitham, G.B.}} \yr{1974} {\em Linear and Nonlinear Waves\/}.
  \publ{Wiley, NY}.

\bibitem[Yih(1959)]{Yih1959}
{\sc \au{Yih, C.-S.}} \yr{1959}  \at{Effects of gravitational or
  electromagnetic fields on fluid motion.}  \jt{{Quarterly of Applied
  Mathematics}}  \bvol{16}~(4),  \pg{409--415}.

\bibitem[Yih(1988)]{Yih1988}
{\sc \au{Yih, C.-S.}} \yr{1988} {\em Fluid Mechanics\/}.  \publ{West River
  Press, Ann Arbor MI}.

\end{thebibliography}
\end{document}